\documentclass[twocolumn,groupedaddress,amsmath,amssymb,aps,pra]{revtex4-2}

\usepackage{natbib}
\usepackage{graphicx}
\usepackage[utf8]{inputenc}
\usepackage{booktabs}
\usepackage{dcolumn}
\usepackage{bm}
\usepackage{physics}
\usepackage{qcircuit}
\usepackage{siunitx}
\usepackage{color}
\usepackage[dvipsnames]{xcolor}
\usepackage{xspace}
\usepackage{float}
\usepackage[colorlinks=true, linkcolor=WildStrawberry, urlcolor=WildStrawberry, citecolor=Emerald, linktocpage=true, breaklinks=true, bookmarksopen=true]{hyperref}
\makeatletter
\let\old@makecaption=\@makecaption
\usepackage{subcaption}
\let\@makecaption=\old@makecaption
\makeatother

\newcommand{\ceil}[1]{\lceil#1\rceil}
\newcommand{\bigO}[1]{\mathcal{O}(#1)}
\newcommand{\uu}{\textsc{$U2$}\xspace}
\newcommand{\cx}{\textsf{cX}\xspace}
\newcommand{\ccx}{\textsf{ccX}\xspace}
\newcommand{\x}{\textsc{$X$}\xspace}
\newcommand{\swap}{\textsf{SWAP}\xspace}

\newcommand{\mtarg}{*+<.02em,.02em>{\xy ="i","i"-<.39em,0em>;"i"+<.39em,0em>
**\dir{-}, "i"-<0em,.39em>;"i"+<0em,.39em>
**\dir{-},"i"*\xycircle<.4em>{}*\xycircle<.5em>{} \endxy} \qw}

\begin{document}

\title{Demonstration of Shor's factoring algorithm for N=21 on IBM quantum processors}

\author{Unathi \surname{Skosana}}
\email{ukskosana@gmail.com}

\author{Mark \surname{Tame}}
\affiliation{Department of Physics, Stellenbosch University, Matieland 7602, South Africa}
\date{\today}

\begin{abstract}
We report a proof-of-concept demonstration of a quantum order-finding algorithm for factoring the integer 21. Our demonstration involves the use of a compiled version of the quantum phase estimation routine, and builds upon a previous demonstration by Mart\'in-L\'{o}pez {\it et al.} in Nature Photonics 6, 773 (2012). We go beyond this work by using a configuration of approximate Toffoli gates with residual phase shifts, which preserves the functional correctness and allows us to achieve a complete factoring of $N=21$. We implemented the algorithm on IBM quantum processors using only 5 qubits and successfully verified the presence of entanglement between the control and work register qubits, which is a necessary condition for the algorithm's speedup in general. The techniques we employ may be useful in carrying out Shor's algorithm for larger integers, or other algorithms in systems with a limited number of noisy qubits.
\end{abstract}

\maketitle


\section{Introduction \label{sec:introduction}}
Shor's algorithm~\cite{Shor_1997} is a quantum algorithm that provides a way of finding the nontrivial factors of an $L$-bit odd composite integer $N=pq$ in polynomial time with high probability. The crux of Shor's algorithm rests upon Quantum Phase Estimation (QPE)~\cite{Mike&Ike}, which is a quantum routine that estimates the phase $\varphi_u$ of an eigenvalue $e^{2\pi i \varphi_u}$ corresponding to an eigenvector $\ket{u}$ for some unitary matrix $\hat{U}$. QPE efficiently solves a problem related to factoring, known as the order-finding problem, in polynomial time in the number of bits needed to specify the problem, which in this case is $L=\ceil{\log_{2}{N}}$. By solving the order-finding problem using QPE and carrying out a few extra steps, one can factor the integer $N$. There is no known classical algorithm that can solve the same problem in polynomial time~\cite{Mike&Ike, dewolf2019}.

A large corpus of work has been done with regards to the experimental realization of Shor's algorithm over the years. The pioneering work was performed with liquid-state nuclear magnetic resonance, factoring $15$ on a $7$-qubit quantum computer~\cite{Nat414883a.10.1038}. The considerable resource demands of Shor's original algorithm were circumvented by using various approaches, including adiabatic quantum computing~\cite{Peng2008} and in the standard network model using techniques of compilation~\cite{PhysRevLett.91.147902} that reduced the demands to within the reach of single-photon architectures~\cite{PhysRevLett.99.250504, PhysRevLett.99.250505, Scie.1173731.10.1126} and a super-conducting phase qubit system~\cite{Nphys2385.10.1038}. In 2012, a proof-of-concept demonstration of the order-finding algorithm for the integer $21$ was carried out with photonic qubits using, in addition to the aforementioned compilation technique, an iterative scheme~\cite{Nphoton.2012.259.10.1038}, where the control register is reduced to one qubit and this qubit is reset and reused~\cite{PhysRevLett.76.3228, Scie.1110335.10.1126}. However, factoring was not possible in this demonstration due to the low number of iterations. Later, the iterative scheme was demonstrated for factoring 15, 21 and 35 on an IBM quantum processor by splitting up the iterations and combining the outcomes~\cite{Amico2019}. Recently, building on previous schemes of hybrid factorization~\cite{Pal_2019, PhysRevLett.108.130501}, a quantum-classical hybrid scheme has been implemented on IBM's quantum processors for the prime factorization of $35$. This hybrid scheme of factorization alleviates the resource requirements of the algorithm at the expense of performing part of the factoring classically~\cite{Saxena_2020}.

In this paper, we build on the order-finding routine of Ref.~\cite{Nphoton.2012.259.10.1038} and implement a version of Shor's algorithm for factoring 21 using only $5$ qubits -- the work register contains 2 qubits and the control register contains $3$ qubits, each providing 1-bit of accuracy in the resolution of the peaks in the output probability distribution used to find the order. This approach is in contrast to the iterative version~\cite{PhysRevLett.85.3049} used in Refs.~\cite{Nphoton.2012.259.10.1038} and~\cite{Amico2019}, which employs a single qubit that is recycled through measurement and feed-forward, giving 1-bit of accuracy each time it is recycled. The advantage of the iterative approach lies in this very reason; through mid-circuit measurement and real-time conditional feed-forward operations, the total number of qubits required by the algorithm is significantly reduced. At the time of writing, IBM's quantum processors do not yet support real-time conditionals necessary for the implementation of the iterative approach, so we use 3 qubits for the control register, one for each effective iteration. Thus, our compact approach is completely equivalent to the iterative approach. In future, once the capability of performing real-time conditionals is added, a further reduction in resources will be possible for our implementation, potentially improving the quality of the results even more and opening up the possibility of factoring larger integers.

As it stands, the controlled-NOT (\cx) gate count of the standard algorithm~\cite{QIC2011517.2011525} exceeds $40$ and in preliminary tests we have found that the output probability distribution is indistinguishable from a uniform probability distribution (noise) on the IBM quantum processors. Our improved version reduces the \cx gate count through the use of relative phase Toffoli gates, reducing the \cx gate count by half while leaving the overall operation of the circuit unchanged and we suspect this technique may extend beyond the case considered here. We have gone further than the work in Ref.~\cite{Nphoton.2012.259.10.1038}, where full factorization of 21 was not achieved as with only two bits of accuracy for the peaks of the output probability distribution continued fractions would fail to extract the correct order. On the other hand, in the work in Ref.~\cite{Amico2019}, where 21 was factored on an IBM processor, a larger number of 6 qubits was required and the iterations were split into three separate circuits, with the need to re-initialise the work register into specific quantum states for each iteration. Our approach is thus more efficient and compact, enabling algorithm outcomes with reduced noise. To support our claims, we successfully carry out continued fractions and evaluate the performance of the algorithm by (i) quantitatively comparing the measured probability distribution with the ideal distribution and noise via the Kolmogorov distance, (ii) performing state tomography experiments on the control register, and (iii) verifying the presence of entanglement across both registers.

The paper is organized as follows. In Sec.~\ref{sec:background}, we give a brief review of the order-finding problem and its relation to Shor's algorithm. In Sec.~\ref{sec:compiled_shors_routine} we expound on the compiled version of Shor's algorithm, where we consider the specific case of the factorization of $N=21$. We construct the quantum circuits that realize the required modular exponentiation unitaries and proceed to optimize their \cx gate count through the introduction of relative phase Toffoli gates. We report our results from executing our compact construction of the algorithm on IBM's quantum computers in Sec.~\ref{sec:exps_on_IBM}. Finally, we provide concluding remarks of our study in Sec.~\ref{sec:conclusion}. An appendix is also included.


\section{Background}\label{sec:background}
\subsection{Order finding}
The order-finding problem is typically stated as follows. Given positive integers $N$ and $a \in \{0, 1, \ldots, N - 1\}$ that share no common factors, we seek to find the least positive integer $r \in \{0, 1, \ldots N\}$ such that $a^{r}\!\mod N = 1$. The integer $r$ is said to be the \emph{order} of $a$ and $N$, and the order-finding problem is that of finding $r$ for a particular $a$ and $N$. There exists no classical algorithm that can solve the order-finding problem efficiently, that is, with operations (elementary gates) that scale polynomially in the number of bits needed to specify $N$, \emph{i.e.} $L \equiv \ceil{\log_{2}{N}}$~\cite{Mike&Ike, dewolf2019}.


\subsection{Shor's algorithm}\label{sec:shors}
The order-finding problem can be efficiently solved on a quantum computer with $\bigO{L^3}$ operations; the cost being mostly due to the \emph{modular exponentiation} operation which requires $\bigO{L^3}$ quantum gates~\cite{Mike&Ike}. The problem of prime factorization is the subject of Shor's algorithm, which is equivalent to the order-finding problem: for an $L$-bit positive odd integer $N=pq$ and randomly chosen positive integer $a \leq N$ co-prime to $N$, the order $r$ of $a$ and $N$ can be used to find the non-trivial factors of $N$. The algorithm is probabilistically guaranteed, with probability greater than a half that the greatest common divisor $\gcd(a^{r/2} \pm 1, N)$ gives the prime factors of $N$~\cite{Mike&Ike}. Shor's algorithm uses two quantum registers; a control register and a work register. The control register contains $n$ qubits, each for one bit of precision in the algorithmic output. The work register contains $m = \ceil{\log_{2}{N}}$ qubits where $m$ is the number of qubits to encode $N$. The measurement of the control register outputs a probability distribution peaked at approximately the values of $2^n s/r$, where $s$ is associated with the outcome of the measurement and thus randomly assigned. The details of how the peaked probability distribution comes about are given in the order-finding routine outlined below. One can determine the order $r$ from the peak values of the distribution using \emph{continued fractions}, with a number of operations that scales polynomially in $\ceil{\log_{2}{N}}$. The procedure, or routine, for order finding is summarized below.


\subsubsection*{Order-finding routine}\label{sec:procedure}

\begin{enumerate}
\item \emph{Initialization}\newline
Prepare $\ket{0}^{\otimes n}\ket{0}^{\otimes m}$ and apply $H^{\otimes n}$ on the control register and $X$ on the $m^\text{th}$ qubit in the work register to create a superposition of $2^n$ states in the control register and $\ket{1}$ in the work register:
\begin{align*}
    \ket{0}^{\otimes n}\ket{0}^{\otimes m}\to \frac{1}{2^{n/2}}\displaystyle\sum_{x=0}^{2^n-1} \ket{x}\ket{1}.
\end{align*}

\begin{figure}[t]
   \centering
    \includegraphics[width=70mm]{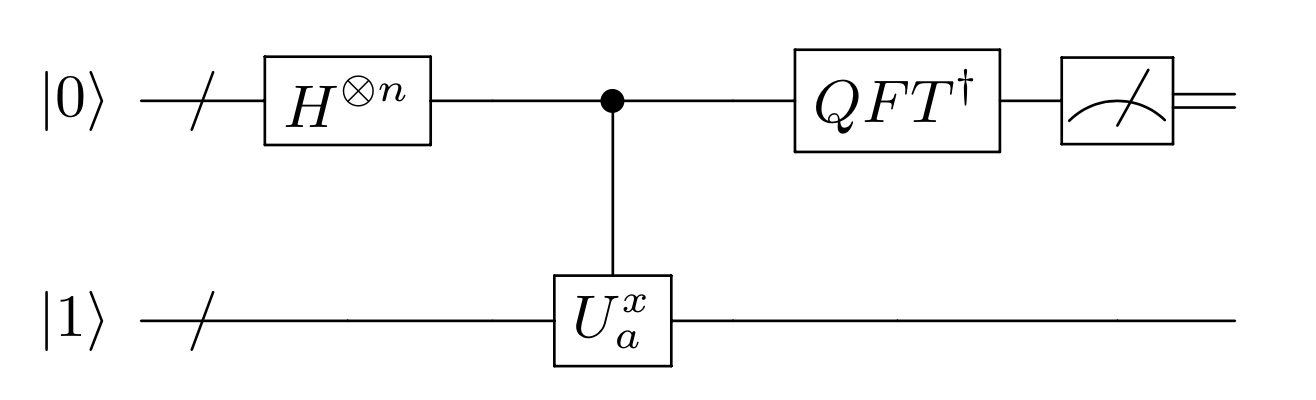} 
        \caption{Schematic of the routine used for the period finding part of Shor's algorithm. The first (control) register has $n$ qubits. The number of qubits in the control register determines the bit-accuracy of the value of $2^ns/r$. The bottom (work) register has the $m$ qubits required to encode $N$. First, the control and work registers are initialized, then conditional modular exponentiation is performed, indicated by the controlled unitary and an inverse quantum Fourier transform is applied to the control register followed by a standard computational basis measurement. The circuit is essentially the QPE algorithm applied to the unitary matrix $\hat{U}_a$ -- see text for details.}
    \label{fig:schematic}
\end{figure}

\item \emph{Modular exponentiation function} (MEF)\newline
Conditionally apply the unitary operation $\hat{U}$ that implements the modular exponentiation function $a^x\>\text{mod}\>N$ on the work register whenever the control register is in state $\ket{x}$:
\begin{align*}
    \frac{1}{2^{n/2}} \displaystyle\sum_{x=0}^{2^n -1} \ket{x}\ket{1} & \to \frac{1}{2^{n/2}}\displaystyle\sum_{x=0}^{2^n-1}
    \ket{x}\ket{a^x\>\text{mod}\>N} \\
    & =
    \frac{1}{\sqrt{r2^n}}\displaystyle\sum_{s=0}^{r-1}\displaystyle\sum_{x=0}^{2^n-1}
    e^{2\pi i s x / r}\ket{x}\ket{u_s}. \\
\end{align*}
In the second line, $\ket{u_s}$ is the eigenstate of $\hat{U}: \hat{U}\ket{u_s} = e^{2\pi i s/r}\ket{u_s}$ and $\frac{1}{\sqrt{r}}\displaystyle\sum_{s=0}^{r-1}\ket{u_s} = \ket{1}$ has been used for the work register. The MEF operation is equivalent to applying $\hat{U}^x$ to the work register when the state $\ket{x}$ is in the control register, as shown in Fig.~\ref{fig:schematic}, with $\hat{U}\ket{y}=\ket{ay\mod N}$ for a given state $\ket{y}$ (the subscript $a$ in $\hat{U}$ is suppressed for notational convenience). This provides an alternative way to write the output state and allows a connection between the MEF operation and the QPE algorithm for the unitary operation $\hat{U}$.

\item \emph{Inverse Quantum Fourier Transform} (QFT)\newline
Apply the inverse quantum Fourier transform on the control register:
\begin{align*}
    \frac{1}{\sqrt{r2^n}}\displaystyle\sum_{s=0}^{r-1}\displaystyle\sum_{x=0}^{2^n-1}
    e^{2\pi i s x / r}\ket{x}\ket{u_s} \to \frac{1}{\sqrt{r}}\displaystyle\sum_{s=0}^{r - 1}\ket{\varphi_s}\ket{u_s}.
\end{align*}

\item \emph{Measurements}\newline
    Measure the control register in the computational basis, yielding peaks in the probability for states where $\varphi_s \simeq 2^n s / r$ due to the inverse QFT. Thus, the outcome of the algorithm is probabilistic, however, there is a high probability of obtaining the location of the $\varphi_s$ peaks after only a few runs. The accuracy of $\varphi_s$ to $2^{n} s / r$ is determined by the number of qubits in the control register.

\item \emph{Continued fractions}\newline
    Apply continued fractions to $\varphi=\varphi_s/2^n$ (the approximation of $s / r$) to extract out $r$ from the convergents (see Appendix G for details).

\end{enumerate}


\section{Compiled Shor's algorithm \label{sec:compiled_shors_routine}}
A full-scale implementation of Shor's algorithm to factor an $L$-bit number would require a quantum circuit with $72L^3$ quantum gates acting on $5L + 1$ qubits for the order-finding routine \cite{PhysRevA.54.1034}, \emph{i.e.} factoring $N=21$ would require $9000$ elementary quantum gates acting on $26$ qubits. The overhead in quantum gates comes from the modular exponentiation function part of the algorithm, while the overhead in qubits comes from the level of accuracy needed to successfully carry out the continued fractions part of the algorithm. Such an overhead obviously puts a full-scale implementation beyond the reach of current devices. However, compilation techniques such as the one described in Ref.~\cite{PhysRevA.54.1034}, bridge this gap and allow for small-scale proof-of-concept demonstrations, where the quantum circuit is tailored around properties of the number to be factored. This significantly simplifies the controlled-operations that realize the MEF operation (see previous section), which is the most resource-intensive part of the order-finding routine. The resource demands of the compiled quantum circuit are significantly reduced, making it suitable for quantum devices with low connectivity.

From Ref.~\cite{Nphoton.2012.259.10.1038}, we extend the compiled quantum order-finding routine for the particular case of factoring $N = 21$ with $a=4$ to accommodate another iteration for better precision in the resolution of the peaks for the value of $2^ns /r$. For the case of $N=21$, other choices of $a$ give $2$, $4$ or $6$ for $r$. The cases for $r=2$ or $r=4$ have been demonstrated for $N=15$~\cite{Nat414883a.10.1038, PhysRevLett.99.250504, PhysRevLett.99.250505, Nphys2385.10.1038, Scie.1173731.10.1126} and would bear a similar circuit structure in the present case. With only three iterations, $r=6$ would be out of reach as continued fractions would fail. For $a=4$ we have $r=3$, which is a choice that does not suffer from the aforementioned reasons. Despite $r$ being an odd integer, the algorithm is successful in finding it from $a=4$. This is the case for certain choices of perfect square $a$ and odd $r$, and $a=4$ and $r=3$ is such a case~\cite{Nphoton.2012.259.10.1038}. 

In contrast to Ref.~\cite{Nphoton.2012.259.10.1038}, our implementation is not iterative and uses three qubits for the control register rather than one qubit recycled on every iteration. The iterative version is based on the recursive phase estimation, made possible by the use of the semi-classical QFT~\cite{PhysRevLett.76.3228}. However, we have used the traditional QFT because mid-circuit measurements with real-time conditionals are not possible yet on IBM's quantum processors. The traditional QFT for $3$ qubits (see~\cite{Mike&Ike} - Box 5.1) that we implemented is equivalent to Fig. 1A and Fig. 1B in Ref.~\cite{Scie.1110335.10.1126}. The latter is the semi-classical QFT that makes possible the implementation of the iterative version of Shor. If mid-circuit measurements with real-time conditionals were possible, the $3$-qubit semi-classical QFT would be possible and may improve the quality of the results we present here through the use of only $1$ qubit for the control register, as in Ref.~\cite{Nphoton.2012.259.10.1038}. IBM has suggested that the behaviour of real-time conditionals can be reproduced through post selection of the mid-circuit measurements. However, in the present case the speed up gained would be lost using this post selection method (see Appendix A).

In Ref.~\cite{Nphoton.2012.259.10.1038}, a step that is unique among the compilation steps of previous demonstrations, and central to their demonstration is mapping the three levels $\ket{1}$, $\ket{4}$ and $\ket{16}$ accessed by the possible $2^L=2^5$ levels of the work-register to only a single qutrit system. In our demonstration we also use this step, however IBM processors consist of qubits and so we represent the work register by 3 basis states from a two-qubit system and discard the fourth basis state as a null state. The states encoding the three possible levels of the work register; $\ket{1}$, $\ket{4}$ and $\ket{16}$ are mapped to $\ket{q_0q_1}$ according to
\begin{align}
    \ket{1}  & \mapsto \ket{\log_{4}1}  =\ket{00}, \nonumber \\
    \ket{4}  & \mapsto \ket{\log_{4}4}  =\ket{01}, \nonumber \\
    \ket{16} & \mapsto \ket{\log_{4}16} =\ket{10}.
\end{align}

Therefore instead of evaluating $4^x\>\text{mod}\>21$ in the work register as described in step $2$ of Sec.~\ref{sec:procedure}, the compiled version of Shor's algorithm effectively evaluates $\log_{4}[4^x\>\text{mod}\>21]$ in its place for $x = 0,1\dots2^3-1$~\cite{PhysRevA.54.1034}, which reduces the size of the work register to $2$ qubits in comparison to the $5$ qubits required in the standard construction. Note the ordering of quantum bits in the work register is $\ket{q} = \ket{q_0}\ket{q_1}$, where the rightmost qubit is associated with the least significant bit. Similarly, with the control register we have $\ket{c} = \ket{c_0}\ket{c_1}\ket{c_2}$. In total the algorithm requires 5 qubits: 3 for the control register and 2 for the work register. Implementing the controlled unitaries $\hat{U}^x$ that perform the modular exponentiation $\ket{x}\ket{y} \to \ket{x}\hat{U}^x\ket{y}=\ket{x}\ket{a^x y\>\text{mod}\>N}$ reduces to effectively swapping around the states $\ket{1}$, $\ket{4}$ and $\ket{16}$ in the work register controlled by the corresponding bit of the integer $x$ in the control register, which is given by $x=c_22^0 + c_12^1 + c_02^2$. In other words, $\hat{U}^x=\hat{U}^{c_02^2}\hat{U}^{c_12^1}\hat{U}^{c_22^0}$. Thus, depending on the control qubit $c_i$, one of the following maps is applied:
\begin{align}
    \hat{U}^{1}: \{ \ket{1} \mapsto \ket{4},  \ket{4} \mapsto \ket{16}, \ket{16} \mapsto \ket{1} \}, \nonumber \\
    \hat{U}^{2}: \{ \ket{1} \mapsto \ket{16}, \ket{4} \mapsto \ket{1},  \ket{16} \mapsto \ket{4} \}, \nonumber \\
    \hat{U}^{4}: \{ \ket{1} \mapsto \ket{4},  \ket{4} \mapsto \ket{16}, \ket{16} \mapsto \ket{1} \}.
\end{align}
The next simplification step comes from the fact that these operations on the work register need not be controlled \swap (Fredkin) gates, they can be as simple as \cx gates, as we show next.


\subsection{Modular exponentiation \label{sec:implementation_of_mod_exp}}
Implementing $\hat{U}^{1}$ on the two-qubit work register is simplified considerably by noting that the states $\ket{4}$ and $\ket{16}$ initially have zero amplitude, and thus the operation $\ket{1} \mapsto \ket{4}$ alone is sufficient. This operation can realized with a \cx gate controlled by $\ket{c_2}$ targeting the second work qubit $\ket{q_1}$. 

{
\small 
\begin{equation}
    \Qcircuit @C=.7em @R=1.5em {
        \lstick{\ket{c_2}}    & \ctrl{1}                 & \qw &&&&  \ctrl{2} & \qw \\
        \lstick{\ket{q_0}}  & \multigate{1}{U^{2^0}}   & \qw &&& = \hspace{0.7cm}  & \qw & \qw \\
        \lstick{\ket{q_1}}  & \ghost{U^{2^0}}          & \qw &&&& \targ & \qw 
    }
    \vspace{0.2cm}
\end{equation}
}

Similarly, the implementation of $\hat{U}^{2}$ can be simplified by noting that the states $\ket{1}$ and $\ket{4}$ are the only non-zero amplitude states in the work register after $\hat{U}^1$ may have been applied, thus prompting us to only consider $\ket{1} \mapsto \ket{16}$ and $\ket{4} \mapsto \ket{1}$. A \cx gate controlled by $\ket{c_1}$ targeting $\ket{q_1}$ followed by a Fredkin gate, swapping $\ket{q_0}$ and $\ket{q_1}$ realizes this simplified $\hat{U}^{2}$. 

{
\small
\begin{equation}
    \Qcircuit @C=.7em @R=1.5em {
         \lstick{\ket{c_1}} & \ctrl{1}                 & \qw &&&&  \ctrl{2} & \qw & \ctrl{1} & \qw & \qw \\
         \lstick{\ket{q_0}} & \multigate{1}{U^{2^1}}   & \qw &&& = \hspace{0.7cm}  & \qw & \targ & \ctrl{1}  & \targ & \qw \\
       \lstick{\ket{q_1}}  & \ghost{U^{2^1}}          & \qw &&&& \targ & \ctrl{-1}  & \targ & \ctrl{-1} & \qw
    }
    \vspace{0.2cm}
\end{equation}
}
In the above, the Fredkin gate has been decomposed into a Toffoli gate (\ccx) and two \cx gates. The subsequent implementation of $\hat{U}^{4}$ admits no simplifications as all the possible states in the work register may have non-zero amplitude at this point. This operation is implemented with a Toffoli and a Fredkin gate with single-qubit \x gates.

{
\small
\begin{equation}
    \Qcircuit @C=.7em @R=1.5em {
         \lstick{\ket{c_0}} & \ctrl{1}                 & \qw &&&& \qw & \ctrl{2} & \qw & \qw & \ctrl{1} & \qw & \qw \\
        \lstick{\ket{q_0}} & \multigate{1}{U^{2^2}}   & \qw &&& = \hspace{0.7cm}  & \qw & \targ & \qw & \targ & \ctrl{1}  & \targ & \qw \\
        \lstick{\ket{q_1}} & \ghost{U^{2^2}}          & \qw &&&& \gate{X} & \ctrl{-1} & \gate{X} & \ctrl{-1}  & \targ & \ctrl{-1} & \qw
    }
    \vspace{0.2cm}
\end{equation}
}

\begin{figure*}[t]
 \centering
    \includegraphics[width=150mm]{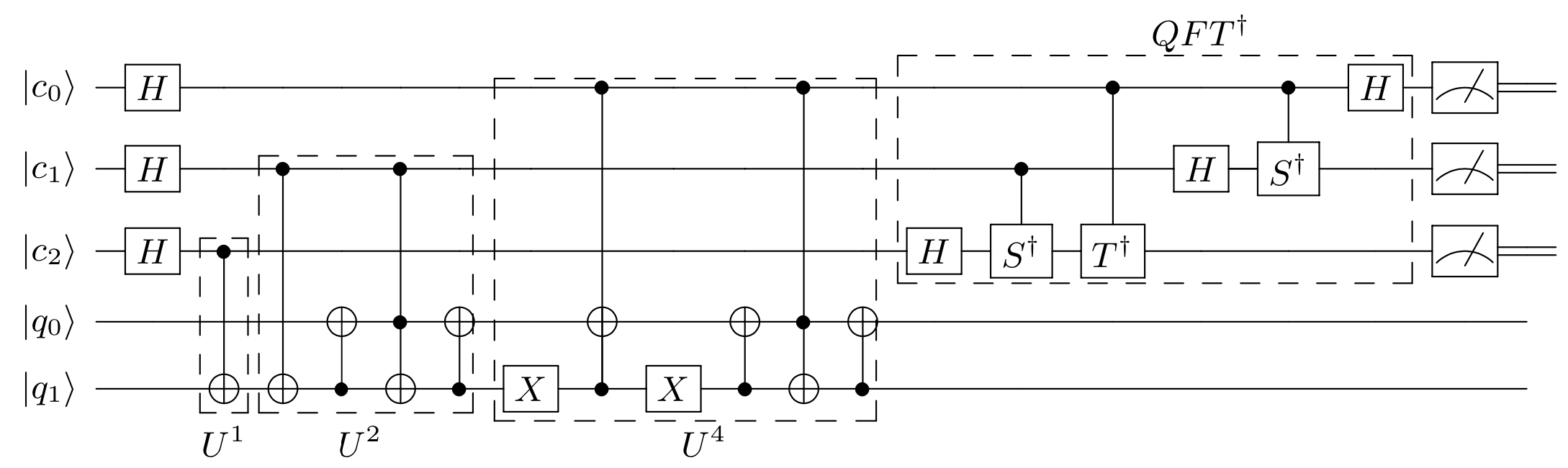}
        \caption{
        Compiled quantum order-finding routine for $N = 21 \text{ and } a = 4$.  This circuit uses five qubits in total; $3$ for the control register
        and $2$ for the work register.  The above circuit determines $2^ns/r$ to three bits of accuracy, from which the order can be extracted. Here, up to a global phase, $S= R_z(\frac{\pi}{2})$ and $T = R_z(\frac{\pi}{4})$ are phase and $\pi/8$ gates, respectively.
    }
    \label{fig:complete}
\end{figure*}

\begin{figure*}[t]
 \centering
    \includegraphics[width=140mm]{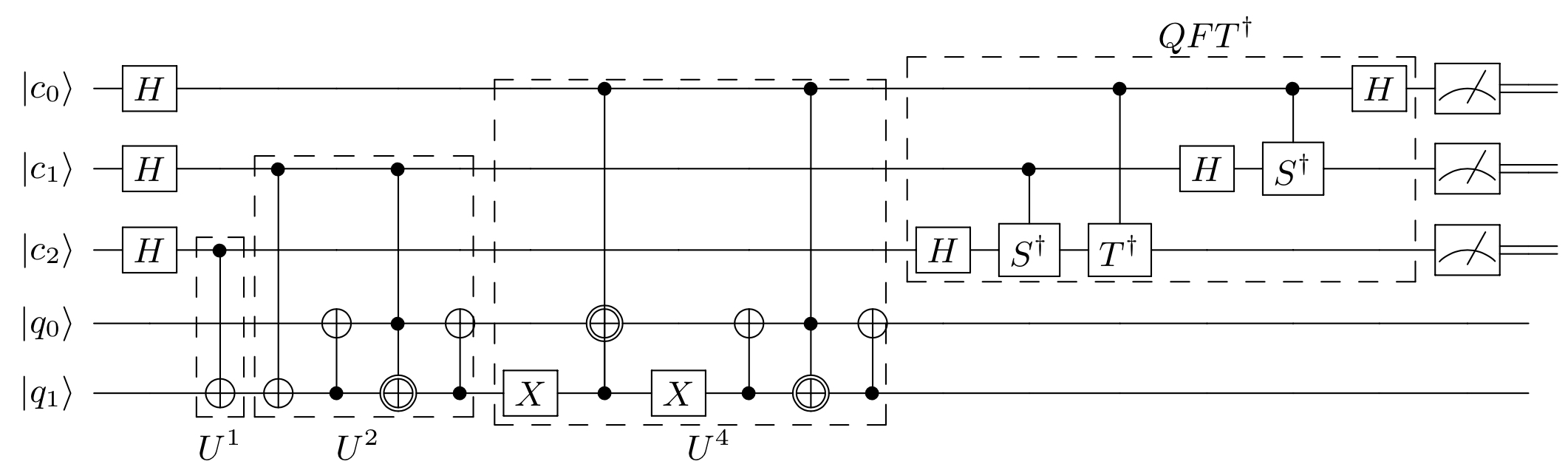}
        \caption{
        Approximate compiled quantum order-finding routine implemented with Margolus gates in place of Toffoli gates in the construction in Fig.~\ref{fig:complete}.
    }
    \label{fig:c_complete}
\end{figure*}

The full circuit diagram is shown in Fig.~\ref{fig:complete} -- note that before simplification the order of application of the controlled unitaries is interchangeable, $\hat{U}^{2^{(n-1)}}$ or $\hat{U}^{2^{1}}$ could be applied first. Interchanging the order only has the effect of interchanging the order of the outcome bits at the end of the computation. This is the reason the order of application of the controlled unitaries here is in reverse order to that in Ref.~\cite{Nphoton.2012.259.10.1038}.


\subsection{Modular exponentiation with relative phase Toffolis \label{sec:mod_exp_with_rel_phase}}
In total, the modular exponentiation routine requires three Toffoli gates; traditionally a single Toffoli gate can be decomposed into six \cx gates and several single-qubit gates~\cite{Mike&Ike} as follows

{
\small
\begin{equation}
    \Qcircuit @C=.5em @R=1.3em {
        & \ctrl{1} & \qw  &&&&  \qw & \qw & \qw & \ctrl{2} & \qw & \qw & \qw 
                             & \ctrl{2} & \qw & \ctrl{1} & \gate{T} & \ctrl{1} & \qw \\ 
        & \ctrl{1} & \qw  &&& = \hspace{0.5cm} &  \qw & \ctrl{1}  & \qw & \qw & \qw 
                             & \ctrl{1}  & \qw       & \qw & \gate{T} & \targ & \gate{T^{\dagger}}
                             & \targ & \qw \\
        & \targ    & \qw &&&& \gate{H} & \targ & \gate{T^{\dagger}} & \targ & \gate{T}
                             & \targ & \gate{T^{\dagger}} & \targ & \gate{T} & \gate{H} & \qw & \qw
                             & \qw \\
    }
    \vspace{0.2cm}
\end{equation}
}

Taking into account a given processor's topology and the constraints it poses, as well as other parts of the circuit (the inverse QFT), further increases the tally of \cx gates. This becomes undesirable as it is understood that there is an upper limit on the number of \cx gates that can be in a circuit with the guarantee of a successful computation. The number of \cx gates from the decomposition of the Toffoli gate can be cut in half if we permit the operation to be correct up to relative phase shifts. Margolus constructed a gate that implements the Toffoli gate up to a relative phase shift of $\ket{101} \mapsto -\ket{101}$ that only uses three \cx gates and four single qubit gates~\cite{Marg_1994}. This construction has been shown to be optimal~\cite{Song_2003}. 

{
\small
\begin{equation}
    \label{eq:margolus}
    \Qcircuit @C=.7em @R=.6em @!R {
        & \ctrl{1}  & \qw &&&&                   \qw      & \qw        & \qw
                             & \ctrl{2} & \qw      & \qw       & \qw & \qw \\
        & \ctrl{1}  & \qw &&& = \hspace{0.5cm} & \qw      & \ctrl{1}   & \qw
                             & \qw      & \qw      & \ctrl{1}  & \qw & \qw \\
        & \mtarg    & \qw &&&& \gate{R_y^{+\frac{\pi}{4}}} & \targ  & \gate{R_y^{+\frac{\pi}{4}}}
                             & \targ &\gate{R_y^{-\frac{\pi}{4}}} & \targ  & \gate{R_y^{-\frac{\pi}{4}}} & \qw 
    }
    \vspace{0.2cm}
\end{equation}
}

The advantages of relative phase Toffoli gates extend beyond the commonly conceived scenarios, {\it i.e.} when the gate is applied last or when the relative phase shifts do not matter for certain configurations of multiply-controlled Toffoli gates. Maslov reported circuit identities that permit the replacement of Toffoli gates with their relative phase variants in certain configurations, resulting in no overall change to the functionality in any significant way~\cite{PhysRevA.93.022311}. The configuration in the circuit shown in Fig.~\ref{fig:complete} is one such configuration that permits a replacement of Toffoli gates with Margolus gates without changing the overall functionality. All the Margolus gates in the circuit in Fig.~\ref{fig:c_complete} (which is the circuit in Fig.~\ref{fig:complete} with the Toffoli gates replaced by Margolus gates) never encounter the basis state $\ket{101}$, thus leaving the operation of the circuit unchanged. See Appendix B for details. This further compacting reduces the number of \cx gates considerably and puts the algorithm within reach of current IBM processors with a limited number of noisy qubits.

\begin{figure*}[t]
   \centering
    \includegraphics[width=130mm]{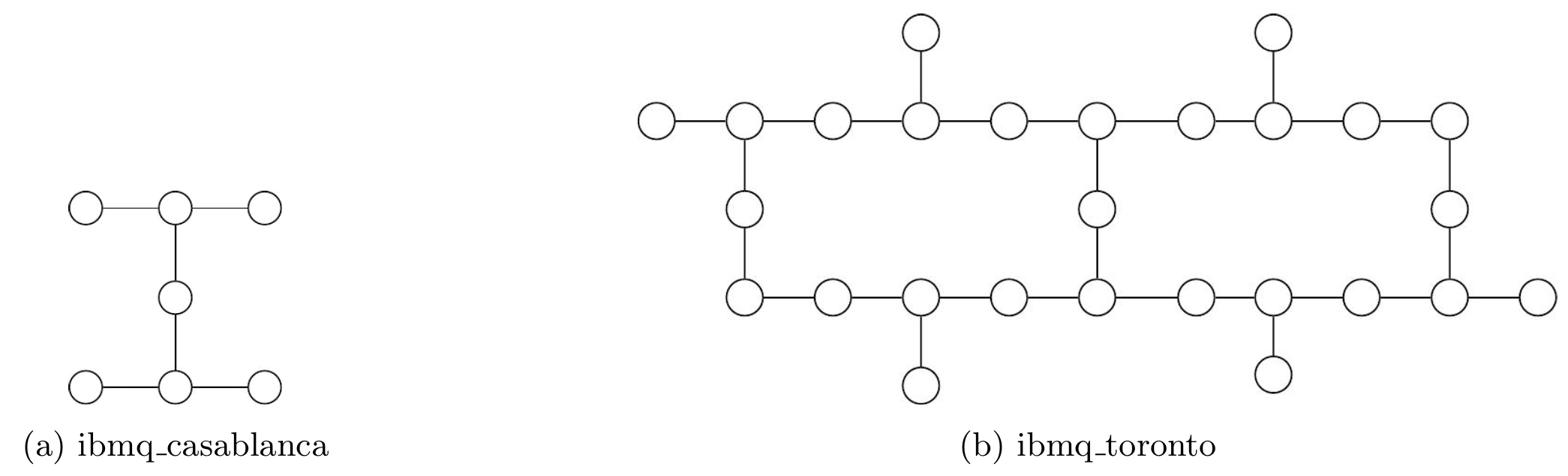}    
    \caption{
        Qubit topology of IBM Q experience processors.
    }
    \label{fig:devices}
\end{figure*}

\begin{figure}[b]
    \centering
    \includegraphics[width=30mm]{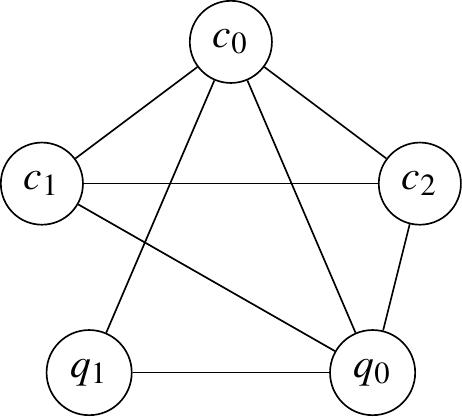}
    \caption{
        Qubit connections required by the compiled circuit in Fig.~\ref{fig:c_complete}.
    }
    \label{fig:mapping}
\end{figure}


\section{Experiments \label{sec:exps_on_IBM}}


\subsection{Physical qubit mapping \label{sec:physical_q_mapping}}
The proposed compiled circuit in Fig.~\ref{fig:c_complete} was mapped onto $5$ physical qubits ($3$ control qubits and $2$ work qubits) and executed on a sub-processor of IBM's 7-qubit quantum processor \textbf{ibmq\_casablanca} and $27$-qubit quantum processor \textbf{ibmq\_toronto}, which we will refer to as 7Q and 27Q, and whose topologies are shown in Fig.~\ref{fig:devices}. When mapping the compiled circuit a few considerations can be taken into account. First, as can be seen from Eq.~\eqref{eq:margolus}, the Margolus gate can be implemented on a collinear set of qubits, as the first control qubit need not be connected to the second control qubit. On the other hand, mapping the three-qubit inverse QFT onto physical qubits without incurring additional \swap gates is not possible, as the three controlled-phase gates require all three qubits to be interconnected in a triangle and the aforementioned quantum processors do not have such a topology. Additionally, more \swap gates are introduced to the transpiled circuit, as the processor topologies do not permit the topology required by the compiled circuit, as shown in Fig.~\ref{fig:mapping}. 

\begin{figure*}[t]
 \centering
    \includegraphics[width=130mm]{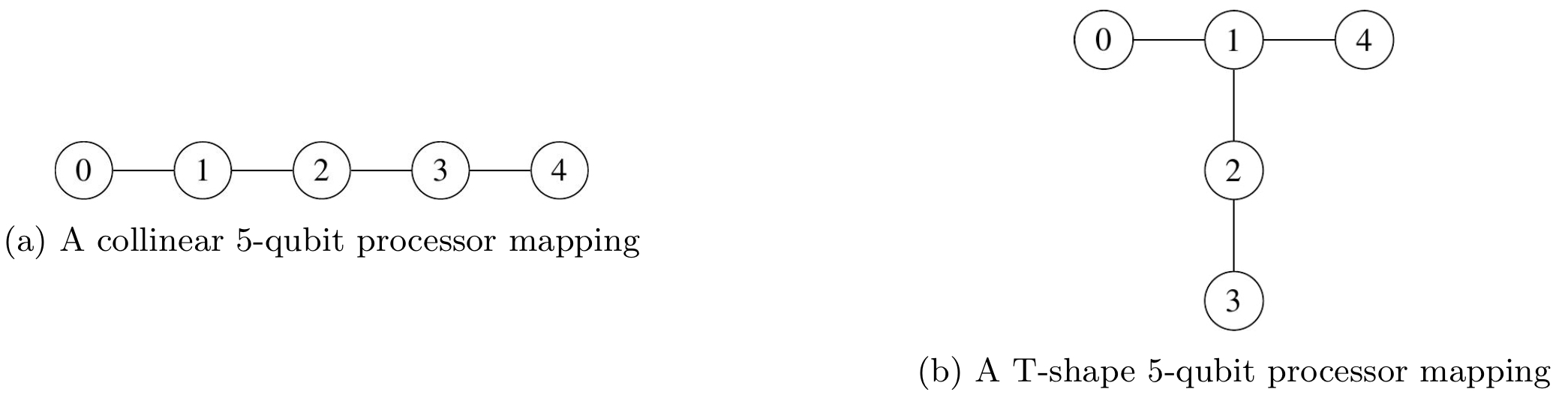}    
    \caption{
        The two possible 5-qubit processor mappings on the architectures shown in Fig.~\ref{fig:devices}.
    }
    \label{fig:shapes}
\end{figure*}

The only possible five-qubit mappings on the quantum processors are all isomorphic to either a collinear set of qubits or a T-shaped set of qubits, as shown in Fig.~\ref{fig:shapes}~a and b. Choosing the mapping in Fig.~\ref{fig:shapes}~b over the one in Fig.~\ref{fig:shapes}~a is motivated by the fact that the former is slightly more connected than latter and thus in effect would reduce the number of \swap gates in the mapped and transpiled circuit.


\subsection{Performance \label{sec:performance}}
\begin{figure}[b!]
    \centering
    \includegraphics[width=66mm]{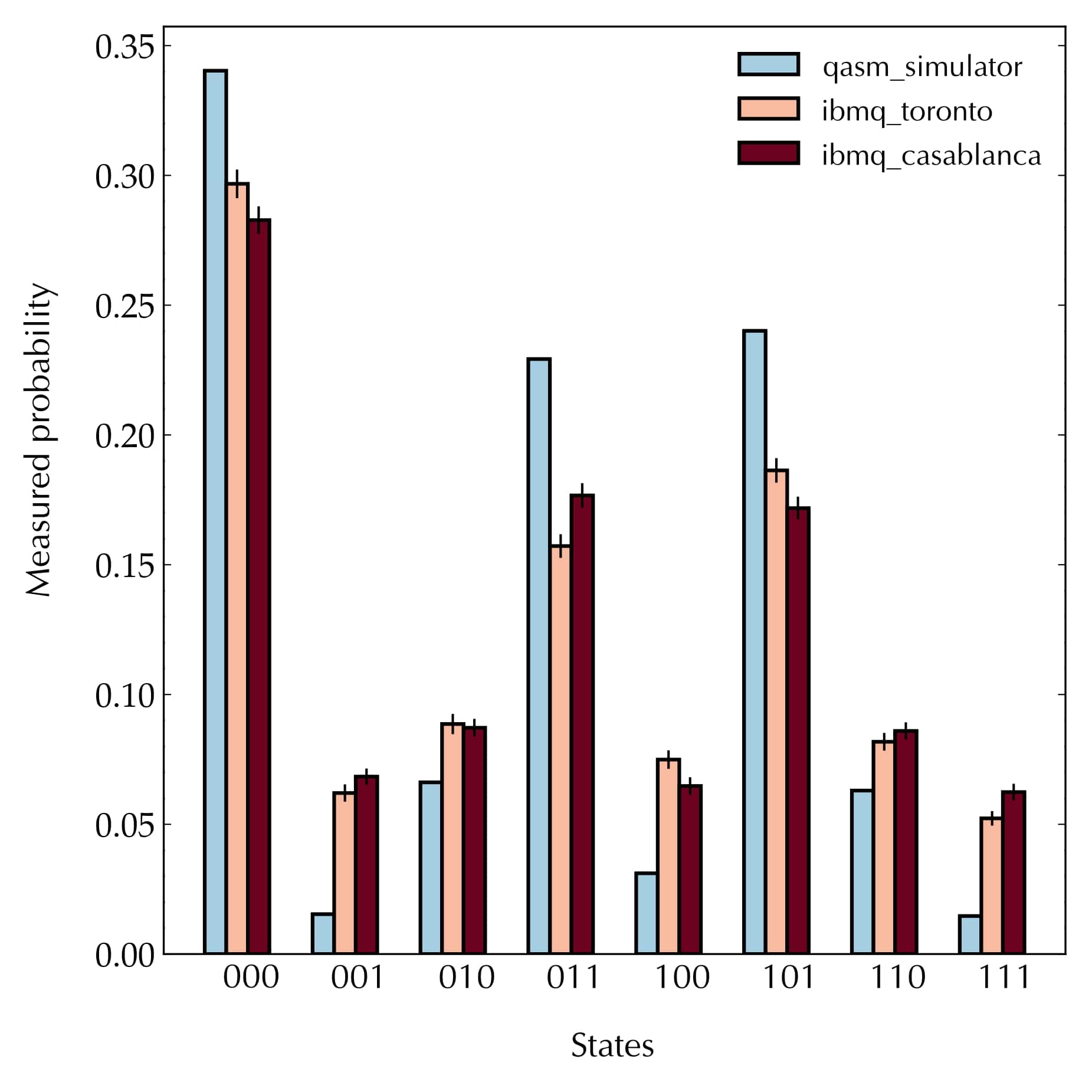}
    \caption{
        Results of the complete quantum order-finding routine for $N = 21$ and $a=4$. On each processor, the circuit was executed $8192 \times 100$
        times with measurement error mitigation. The error bars represent $95\%$ confidence intervals around the mean value of each histogram bin (see Appendix D). The simulator probabilities show the ideal case.
    }
    \label{fig:histogram}
\end{figure}

To evaluate the performance of the algorithm, we first transpiled the circuit in Fig.~\ref{fig:c_complete} down to the chosen quantum processor with the mapping below
\begin{align}
    0 \mapsto c_0, \nonumber \\
    1 \mapsto c_2, \nonumber \\
    4 \mapsto c_1, \nonumber \\
    2 \mapsto q_1, \nonumber \\
    3 \mapsto q_0.
\end{align}
Through the transpiler's optimization, with the mapping above it is possible to have a circuit that has $25$ \cx gates and a circuit depth of $35$. Fig.~\ref{fig:histogram} shows the results of measurements on the control register qubits from the two processors, where measurement error mitigation has been applied to results and mitigates the effect of measurement errors on the raw results (see Appendix C). The outcomes $\ket{011}$ and $\ket{101}$ occur with probability $\sim 16\%$ and $\sim 19\%$ on \textbf{ibmq\_toronto} and $\sim 18\%$ and $\sim 17\%$ on \textbf{ibmq\_casablanca}, respectively. The theoretical ideal probability is $\sim25\%$, as can be seen from the simulator results in Fig.~\ref{fig:histogram}. However, the amplification of the peaks $\ket{000}$, $\ket{011}$ and $\ket{101}$ is clearly visible from the processor outcomes. 

\begin{figure}[hbt]
    \centering
    \includegraphics[width=66mm]{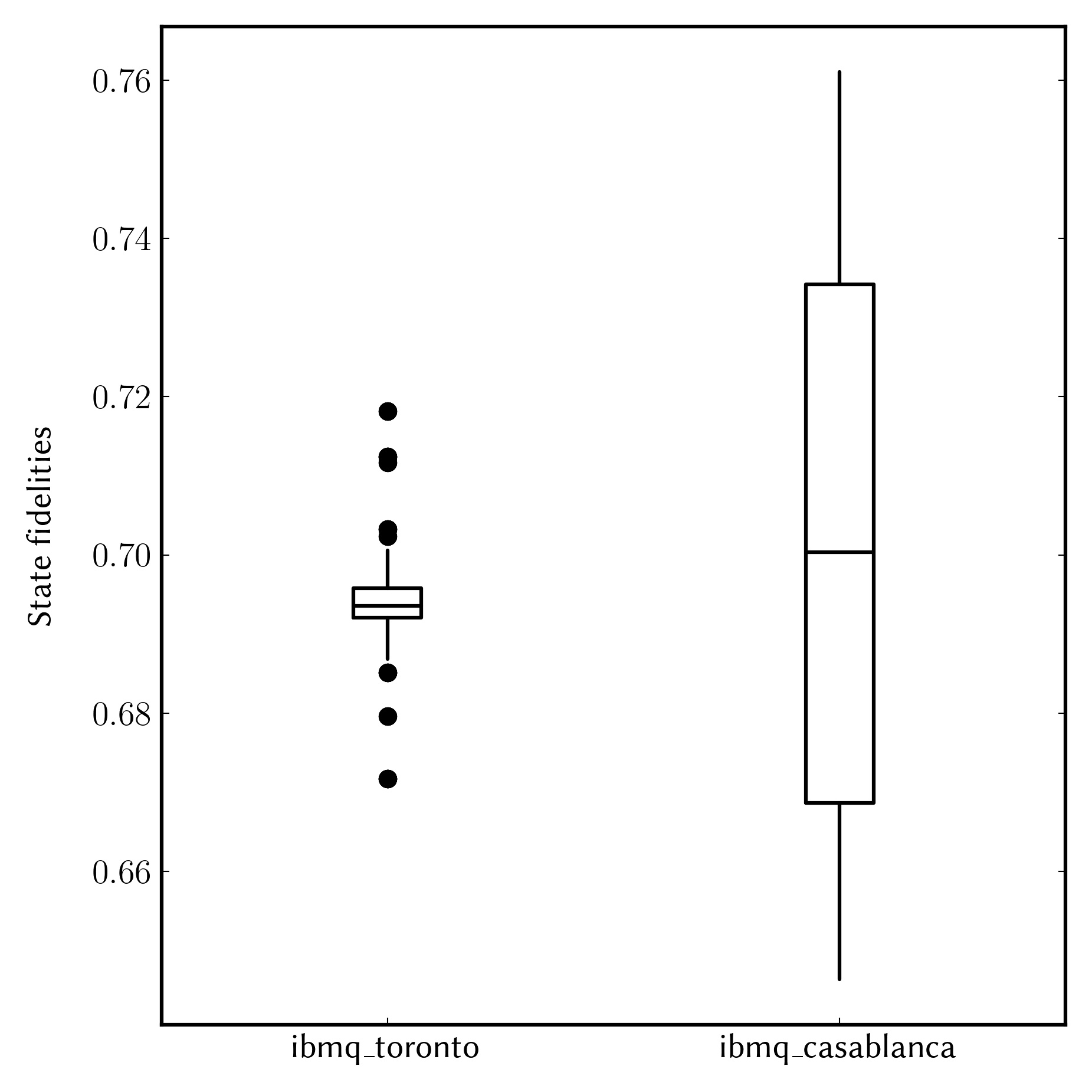}
    \caption{
        Boxplot of a sample ($\nu = 50$) of state fidelities from the respective two devices showing the spread of the values around the sample mean and
        $95\%$ confidence intervals.
    }
    \label{fig:box_plot}
\end{figure}

\begin{figure*}[t]
    \centering
    \includegraphics[width=170mm]{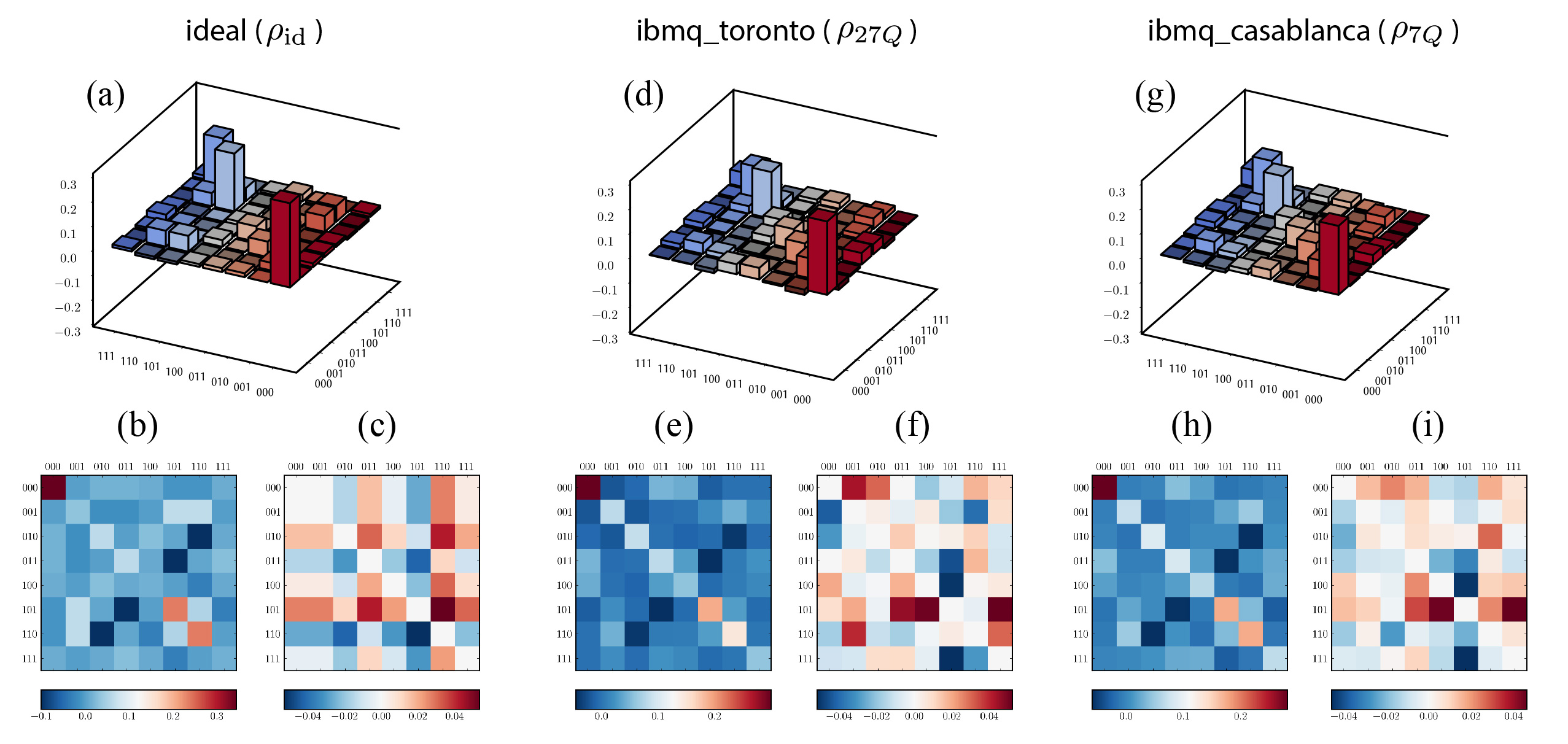}
    \caption{
        Ideal and measured density matrices after the inverse QFT, estimated via a maximum-likelihood reconstruction from measurement results in the Pauli-basis. (a) The ideal state $\op{\Psi}$ (only the real parts are shown, imaginary parts are less than $0.04$). (b) A matrix plot of the real part of $\op{\Psi}$. (c) A matrix plot of the imaginary part of $\op{\Psi}$. These plots are compared with the measured states $\rho_{27Q}$ and $\rho_{7Q}$ in panels (d) and (g), and the corresponding matrix plot of their real parts in panels (e) and (h), and imaginary parts in panels (f) and (i), respectively. We observe there is a resemblance between the ideal state and the measured states, but noise in both real and imaginary parts is notable.
    }
    \label{fig:density_mats}
\end{figure*}

We quantify the successful performance of the algorithm by comparing the experimental and ideal probability distributions via the trace distance or Kolmogorov distance~\cite{Mike&Ike}, which measures the closeness of two discrete probability distributions $P$ and $Q$ and is defined by the equation $D(P,Q) \equiv \sum_{x \in \mathcal{X}}|P(x) - Q(x)|/2$, where ${\cal X}$ represents all possible outcomes. This measure shows an agreement between measured and ideal results -- the trace distance between the measured distribution and the ideal distribution is $0.1694$ and $0.1784$ for \textbf{ibmq\_toronto} and \textbf{ibmq\_casablanca}, respectively. On the other hand, the trace distance between the ideal distribution and a candidate random uniform distribution is $0.4347$. Furthermore, we evaluate the performance of the algorithm by characterizing the measured output state in the control register, this is achieved via state tomography yielding the density matrix of the measured state. The measured state and ideal state on the output register are quantitatively compared using the fidelity for two quantum states $\rho$ and $\sigma$, and is defined to be $F(\rho, \sigma) \equiv \tr\sqrt{\rho^{1/2}\sigma\rho^{1/2}}$~\cite{Mike&Ike}. We measured a fidelity of $F(\rho_\text{id}, \rho_{27Q})=0.6948 \pm 00650$ and $F(\rho_\text{id}, \rho_{7Q}) = 0.70 \pm 0.0275$ on the $27$ qubit and $7$ qubit quantum processors respectively, as shown in Fig.~\ref{fig:box_plot}. In Fig.~\ref{fig:density_mats} we show the estimated density matrices in the computational basis for each respective device.

\subsection{Factoring $N=21$ \label{sec:factoring}}

The measured probability distributions in Fig.~\ref{fig:histogram} are peaked in probability for the outcomes $000~(\varphi_s=0)$, $011~(\varphi_s=3)$ and $101~(\varphi_s=5)$, with ideal probabilities of $0.35$, $0.25$ and $0.25$, respectively. Here we are using the integer representation of the binary outcomes. The outcome $000$ corresponds to a failure of the algorithm~\cite{Nphoton.2012.259.10.1038}. For the outcome $011$, computing the continued fraction expansion of $\varphi=\varphi_s/2^n=3/2^3=3/8$ gives the convergents $\{0, 1/2, 1/3, 3/8\}$ (see Appendix G for details), so that the third convergent $1/3$ in the expansion can be identified as $s/r$ and correctly gives $r=3$ as the order when tested with the relation $a^r {\rm mod}\,N=1$, while the other convergents do not give an $r$ that passes the test. Also, for $101$, computing the continued fraction expansion of $\varphi=\varphi_s/2^n=5/2^3=5/8$ gives the convergents $\{0, 1, 1/2, 2/3, 5/8\}$ (see Appendix G for details), so that the third convergent $2/3$ in the expansion can be identified as $s/r$ and correctly gives $r=3$ as the order, while the other convergents do not give an $r$ that passes the test. 

On the other hand, adjacent outcomes that have an appreciable but lower probability do not give the correct order, for example for the outcome $110$ the continued fraction expansion of $\varphi=6/8$ gives $\{0, 1, 3/4\}$ and incorrectly gives $r=4$ as the order (see Appendix G for details). If the peaks for the outcomes are not well distinguished after a fixed number of shots, this type of failure in identifying the order can be avoided in principle by adding further qubits to the control register so that the peak in the probability distribution becomes narrower and more well defined~\cite{Nphoton.2012.259.10.1038}. It is interesting to note that from the results of Ref.~\cite{Nphoton.2012.259.10.1038}, successfully finding the order $r=3$ was not possible to achieve, as with only two bits of accuracy in the experiment the continued fractions would always fail due to the peaked outcomes of $10~(2)$ and $11~(3)$ giving the convergents of $\{0,1/2\}$ and $\{0,1,3/4\}$, respectively. In our case, we successfully find $r=3$, from which we obtain ${\rm gcd}(a^{r/2}\pm1,N)={\rm gcd}(8\pm1,21)=3$ and 7. Thus, with our demonstration, extending the number of outcome bits to three has allowed us to fully perform the quantum factoring of $N=21$.


\subsection{Verification of entanglement \label{sec:verification_of_ent}}
The presence of entanglement between the control and work registers is known to be a requirement for the algorithm to gain any advantageous speedup over its classical counterparts in general~\cite{PhysRevLett.91.147902, Braunstein_1999, Jozsa_2003}. For detecting genuine multipartite entanglement around the vicinity of an ideal state $\ket{\psi}$, one can construct a projector-based witness such as the one below:
\begin{align}
    \hat{\mathcal{W}}_{\psi} = \alpha \mathbb{I} - \op{\psi}{\psi},
\end{align}
where $\alpha$ is the square of the maximum overlap between $\ket{\psi}$ and all biseparable states. In other words, $\tr(\hat{\mathcal{W}}_{\psi}\rho) \geq 0$ for biseparable states and $\tr(\hat{\mathcal{W}}_{\psi}\rho) < 0$ for states with genuine multipartite entanglement in the vicinity of $\ket{\psi}$~\cite{PhysRevLett.92.087902}. For the ideal state after modular exponentiation (but before the inverse QFT) in both the control and work registers, $\alpha=0.75$ was found using the method described in the appendix of Ref.~\cite{PhysRevLett.92.087902}. This was implemented using the software package QUBIT4MATLAB~\cite{TOTH2008430}. Therefore ideally the state in both registers after modular exponentiation has genuine multipartite entanglement.

\begin{figure*}[t!]
    \centering
    \includegraphics[width=170mm]{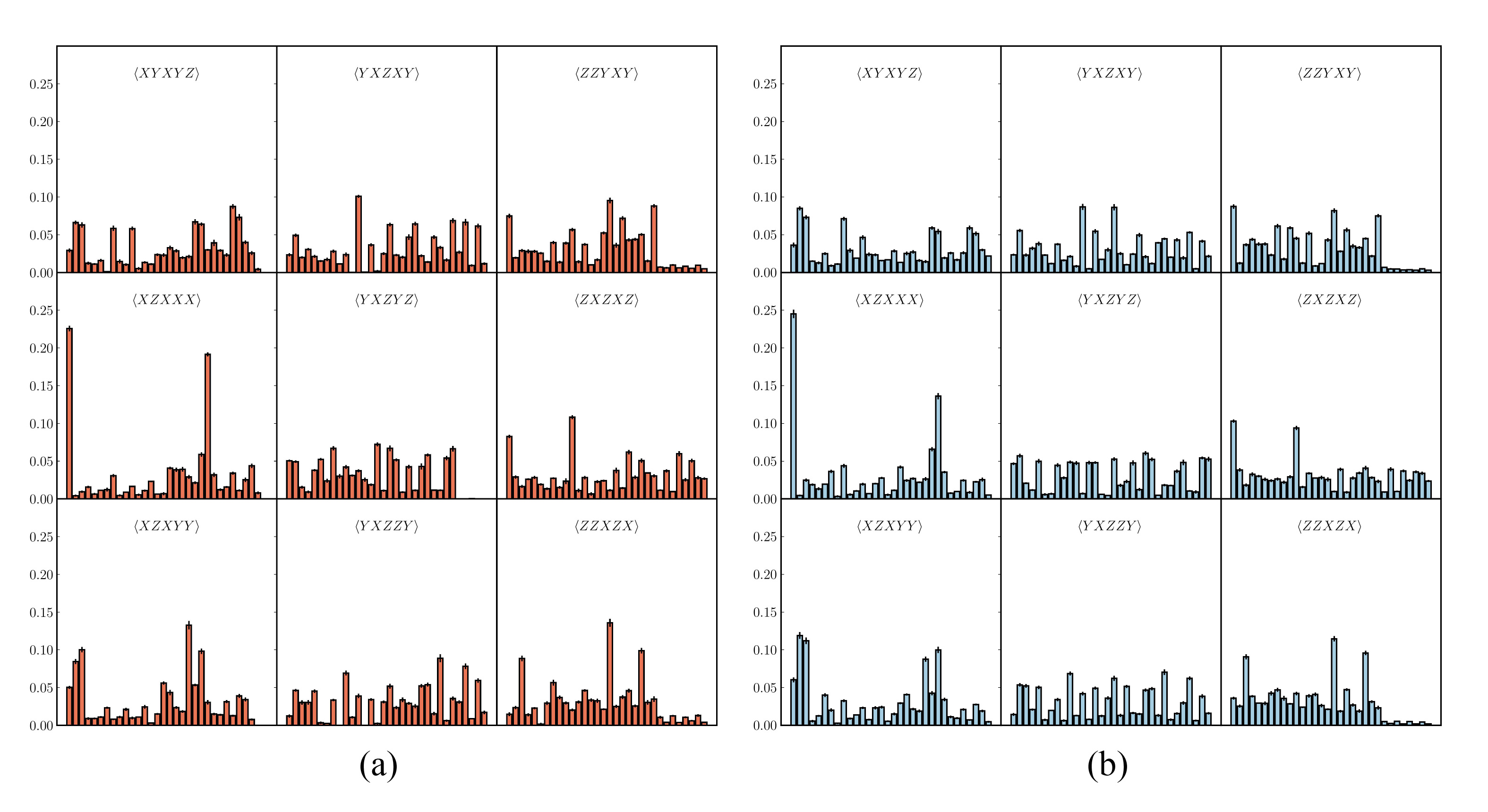}    
    \caption{
        A subset of 9 of the 79 measurement settings required for each term in: (a) $\tr(\op{\Psi}{\Psi}\rho_{7Q})$ and 
        (b) $\tr(\op{\Psi}{\Psi}\rho_{27Q})$. The $x$-axis from left to right shows the labels from $p_{00000}$ to $p_{11111}$.
    }
    \label{fig:klocals}
\end{figure*}

In order to check whether the output state from the IBM processors is close to the ideal state and has genuine multipartite entanglement, full state tomography would normally be needed to characterize the state $\rho_\text{exp}$ in both the control and work registers. This would require $3^5$ measurements, making it impractical to gather a sufficiently large data set within a meaningful time frame. However, we need not measure the full density matrix, the quantity $\tr(\op{\Psi}{\Psi}\rho_\text{exp})$ suffices. To measure this, we can decompose $\rho=\op{\Psi}{\Psi}$ into $293$ Pauli expectations as
\begin{align}
    \op{\Psi}{\Psi} = \displaystyle\sum_{ijklm} p_{ijklm} \sigma_i^{(1)}\sigma_j^{(2)}\sigma_k^{(3)}\sigma_l^{(4)}\sigma_m^{(5)},
\end{align}
where $\sigma_{i} = \{I, X, Y, Z\}$ are the usual Pauli matrices plus the identity. However, the number of measurements needed to obtain all 293 expectation values can be reduced~\cite{PhysRevA.93.032140}. This is because the measured probabilities from a measurement of a single Pauli expectation value, \emph{i.e.} $\expval{ZZZZZ}$, can be summed in various combinations to derive other Pauli expectations values, \emph{i.e.} $\expval{ZIZZZ}, \expval{IZZZZ}$, etc. The values derived are nothing but the marginalization of the measured probabilities over the outcome space of some set of qubits (see Appendix E for details). We can do the same for each term in the set of terms from the Pauli decomposition of $\rho$, calling it $\mathcal{S}_d$, forming a set of other Pauli terms that can be derived from the same counts. Taking the union of these sets to be $\mathcal{S}_u$, the complement $\mathcal{S}_d\backslash\mathcal{S}_u$ gives the $79$ terms we only need to measure (see Appendix E). We measure the 79 Pauli expectation values of the terms above with respect to the state in both registers after modular exponentiation and from this we compute/derive the $293$ terms in $\mathcal{S}_d$ and therefore $\tr(\op{\Psi}{\Psi}\rho_\text{exp})$. The measured probabilities for each term, some of them shown in Fig.~\ref{fig:klocals}, result in an expectation value of $\tr(\op{\Psi}{\Psi}\rho_{7Q}) = 0.677 \pm 0.00365$ and $\tr(\op{\Psi}{\Psi}\rho_{27Q}) = 0.626 \pm 0.00304$, which leads to
\begin{align}
  \tr(\hat{\mathcal{W}}_{\Psi}\rho_{7Q})  &= 0.0729 \pm 0.00365,\nonumber \\
  \tr(\hat{\mathcal{W}}_{\Psi}\rho_{27Q}) &= 0.124 \pm 0.00304.
\end{align}

The results obviously fail to detect genuine multipartite entanglement, however, this does not mean entanglement is entirely absent. Consider the square of the maximum overlap between the ideal state $\ket{\Psi}$ and all pure states $\ket{\theta}$ that are unentangled product states with respect to some bipartite partition (bipartition) $\mathcal{B}$ of the qubits,
\begin{align}\label{eq:maximal_overlap}
    \underset{\theta \in \mathcal{B}}{\max}{\abs{\ip{\theta}{\Psi}}^2} = \beta_{\Psi}.
\end{align}
Thus, any other state $\ket{\xi}$ for which 
\begin{align}
    \abs{\ip{\xi}{\Psi}}^2 > \beta_{\Psi}
\end{align}
cannot be a product state with respect to the bipartition $\mathcal{B}$, implying that there is non-separability, or entanglement, across this bipartition. The above result extends to mixed states $\rho_\xi$ due to the convex sum nature of mixed quantum states~\cite{TOTH2008430}. We compute Eq.~\eqref{eq:maximal_overlap} for all possible bipartitions of our ideal state $\ket{\Psi}$ (see Appendix F for more details). 

For the experimental state $\rho_{7Q}$ we find, with the exception of the bipartition ${\cal B}=(c_0c_1c_2q_1)(q_0)$, that it is non-separable with respect to all other bipartitions, \emph{i.e.} the square of the overlap between $\rho_{7Q}$ and $\ket{\Psi}$ ($\sim 0.677$) is greater than the maximal square overlap between $\ket{\Psi}$ and all product states in each of these bipartitions. Similarly for $\rho_{27Q}$, with the exception of bipartitions ${\cal B}=(c_0c_1c_2q_1)(q_0)$ and ${\cal B}=(c_0c_1c_2q_0)(q_1)$, the state is non-separable with respect to all other bipartitions. Most notably, both $\rho_{7Q}$ and $\rho_{27Q}$ are non-separable with respect to the bipartition ${\cal B}=(c_0c_1c_2)(q_0q_1)$, which is a bipartition between the control and work registers. This implies that non-separability or entanglement is present between the registers, as required for the algorithm's speedup in general~\cite{PhysRevLett.91.147902, Braunstein_1999, Jozsa_2003}. Furthermore, the maximum (not necessarily global but a good proxy of it) expectation value of the operator $\op{\Psi}{\Psi}$ for product states, is found via a greedy search algorithm~\cite{TOTH2008430} to be around $0.30$, further asserting that indeed the qubits are entangled with each other in some way.


\section{Concluding remarks \label{sec:conclusion}}
In summary, we have implemented a compiled version of Shor's algorithm on IBM's quantum processors for the prime factorization of $21$. By using relative phase shift Toffoli gates, we were able to reduce the resource demands that would have been required in the standard compiled and non-iterative construction of Shor's algorithm (with regular Toffoli gates), and still preserve its functional correctness. The use of relative phase shift Toffoli gates has also allowed us to extend the implementation in Ref.~\cite{Nphoton.2012.259.10.1038} to an increased resolution. Moreover, while the latter implementation used only 1 recycled qubit for the control register, in contrast to our 3 qubits, it falls one iteration short of achieving full factoring for the reasons already mentioned. It is not clear what additional resource overheads (single and two-qubit gates) would be needed in implementing another iteration in their scheme and it is likely that these overheads are what prevented the full factoring of 21 in the photonic setup used. Furthermore, we note that in principle there is no real advantage in using 3 qubits for the control register as we have done here instead of 1 qubit recycled, as in Ref.~\cite{Nphoton.2012.259.10.1038}. However, in practice it is not possible at present to recycle qubits on the IBM processors and so we used 3 qubits instead. In future, once this capability is added, a further reduction in resources will be possible for our implementation, potentially improving the quality of the results even more. 

We have verified, via state tomography, the output state in the control register for the algorithm, achieving a fidelity of around $0.70$. For the verification of entanglement generated during the algorithm's operation, the resource demands of state tomography were circumvented by measuring a much reduced number of Pauli measurements to uniquely identify a quantum state~\cite{PhysRevA.93.032140}. However, this method is quite specialized and cannot be easily generalized to larger systems. In scaling up Shor's algorithm to higher integers beyond 21 using larger quantum systems, other methods of quantum tomography can be used to characterize the performance. These include compressed sensing~\cite{PhysRevLett.105.150401} and classical shadows~\cite{Nphyss41567-020-0932-7}, which give theoretical guarantees, and improved scaling in the number of Pauli measurements and classical post-processing than standard methods. In the case when the state belongs to a class of states with certain symmetries, such as stabilizer states, only a few measurements are required for measuring the fidelity and detecting multipartite entanglement~\cite{PhysRevA.72.022340}. However, not all entangled states are neatly housed within these well-studied classes. Ref.~\cite{PhysRevX.7.021042} introduces a device-independent method for multipartite entanglement detection which scales polynomially with the system size by relaxing some constraints. Another scheme constructs witnesses that require a constant number of measurements of the system size at the cost of robustness against white noise. This provides a fast and simple procedure for entanglement detection~\cite{PhysRevLett.117.210504}. Many fundamental questions on the subjects of quantum tomography and multipartite entanglement still remain to be answered~\cite{Banaszek_2013} and advances will help in efficiently quantifying the performance of algorithms in larger quantum processors.

Our demonstration involves a two-fold reduction of the resource count from the full circuit in Fig.~\ref{fig:complete} via the replacement of regular Toffoli gates with relative phase variants, which is an approach that is in the spirit of the NISQ era; tailoring quantum circuits to circumvent the shortcomings of noisy quantum processors. In addition, we suspect that we can further reduce the resource count through the use of the approximate QFT~\cite{PhysRevA.54.139}, while still maintaining a clear resolution of the peaks in the output probability distribution. A possible avenue of future research derived from what we have reported here is the investigation and identification of scenarios where one can replace Toffoli gates with relative phase Toffoli gates while preserving the functional correctness, in a wide range of algorithms including Shor's algorithm, as seen here. In the present case, whether such an approach is special to the case of $N=21$ or extendable to other $N$ is not known. Ref.~\cite{PhysRevA.93.022311} has performed some work in this regard, however a proper analysis and systematic composition of relative phase Toffoli gates for such purposes is still an open problem. In future, a similar approach may make possible the factorization of larger numbers with adequate accuracy in resolution of the algorithm's outcomes and their characterization.

\phantom{.}
\newline
\acknowledgements

We acknowledge the use of IBM Quantum services for this work. The views expressed are those of the authors, and do not reflect the official policy or position of IBM or the IBM Quantum team. We thank Taariq Surtee and Barry Dwolatzky at the University of the Witwatersrand and Ismail Akhalwaya at IBM Research Africa for access to the IBM processors through the Q Network and African Research Universities Alliance. This research was supported by the South African National Research Foundation, the South African Council for Scientific and Industrial Research, and the South African Research Chair Initiative of the Department of Science and Technology and National Research Foundation. We thank David Davis for pointing out an error in the ordering of bits in a previous version.



\begin{thebibliography}{10}
\expandafter\ifx\csname url\endcsname\relax
  \def\url#1{\texttt{#1}}\fi
\expandafter\ifx\csname urlprefix\endcsname\relax\def\urlprefix{URL }\fi
\providecommand{\bibinfo}[2]{#2}
\providecommand{\eprint}[2][]{\url{#2}}

\bibitem{Shor_1997}
\bibinfo{author}{Shor, P.~W.}
\newblock \bibinfo{title}{Polynomial-time algorithms for prime factorization
  and discrete logarithms on a quantum computer}.
\newblock \emph{\bibinfo{journal}{SIAM Journal on Computing}}
  \textbf{\bibinfo{volume}{26}}, \bibinfo{pages}{1484-1509}
  (\bibinfo{year}{1997}).

\bibitem{Mike&Ike}
\bibinfo{author}{Nielsen, M.~A.} \& \bibinfo{author}{Chuang, I.~L.}
\newblock \emph{\bibinfo{title}{Quantum Computation and Quantum Information:
  10th Anniversary Edition}} (\bibinfo{publisher}{Cambridge University Press},
  \bibinfo{address}{USA}, \bibinfo{year}{2011}), \bibinfo{edition}{10th} edn.

\bibitem{dewolf2019}
\bibinfo{author}{de~Wolf, R.}
\newblock \bibinfo{title}{Quantum computing: Lecture notes}
  (\bibinfo{year}{2019}).
\newblock \eprint{1907.09415}.

\bibitem{Nat414883a.10.1038}
\bibinfo{author}{Vandersypen, L. M.~K.} \emph{et~al.}
\newblock \bibinfo{title}{Experimental realization of Shor's quantum
  factoring algorithm using nuclear magnetic resonance}.
\newblock \emph{\bibinfo{journal}{Nature}} \textbf{\bibinfo{volume}{414}},
  \bibinfo{pages}{883-887} (\bibinfo{year}{2001}).

\bibitem{Peng2008}
\bibinfo{author}{Peng, X.} \emph{et~al.}
\newblock \bibinfo{title}{A quantum adiabatic algorithm for factorization and
  its experimental implementation}.
\newblock \emph{\bibinfo{journal}{Phys. Rev. Lett.}}
  \textbf{\bibinfo{volume}{101}}, \bibinfo{pages}{220405}
  (\bibinfo{year}{2008}).

\bibitem{PhysRevLett.91.147902}
\bibinfo{author}{Vidal, G.}
\newblock \bibinfo{title}{Efficient classical simulation of slightly entangled
  quantum computations}.
\newblock \emph{\bibinfo{journal}{Phys. Rev. Lett.}}
  \textbf{\bibinfo{volume}{91}}, \bibinfo{pages}{147902}
  (\bibinfo{year}{2003}).

\bibitem{PhysRevLett.99.250504}
\bibinfo{author}{Lu, C.-Y.}, \bibinfo{author}{Browne, D.~E.},
  \bibinfo{author}{Yang, T.} \& \bibinfo{author}{Pan, J.-W.}
\newblock \bibinfo{title}{Demonstration of a compiled version of Shor's quantum
  factoring algorithm using photonic qubits}.
\newblock \emph{\bibinfo{journal}{Phys. Rev. Lett.}}
  \textbf{\bibinfo{volume}{99}}, \bibinfo{pages}{250504}
  (\bibinfo{year}{2007}).

\bibitem{PhysRevLett.99.250505}
\bibinfo{author}{Lanyon, B.~P.} \emph{et~al.}
\newblock \bibinfo{title}{Experimental demonstration of a compiled version of
  Shor's algorithm with quantum entanglement}.
\newblock \emph{\bibinfo{journal}{Phys. Rev. Lett.}}
  \textbf{\bibinfo{volume}{99}}, \bibinfo{pages}{250505}
  (\bibinfo{year}{2007}).

\bibitem{Scie.1173731.10.1126}
\bibinfo{author}{Politi, A.}, \bibinfo{author}{Matthews, J. C.~F.} \&
  \bibinfo{author}{O'Brien, J.~L.}
\newblock \bibinfo{title}{Shor's quantum factoring algorithm on a photonic
  chip}.
\newblock \emph{\bibinfo{journal}{Science}} \textbf{\bibinfo{volume}{325}},
  \bibinfo{pages}{1221-1221} (\bibinfo{year}{2009}).

\bibitem{Nphys2385.10.1038}
\bibinfo{author}{Lucero, E.} \emph{et~al.}
\newblock \bibinfo{title}{Computing prime factors with a josephson phase qubit
  quantum processor}.
\newblock \emph{\bibinfo{journal}{Nature Physics}}
  \textbf{\bibinfo{volume}{8}}, \bibinfo{pages}{719-723}
  (\bibinfo{year}{2012}).

\bibitem{Nphoton.2012.259.10.1038}
\bibinfo{author}{Mart\'in-L\'opez, E.} \emph{et~al.}
\newblock \bibinfo{title}{Experimental realization of Shor's quantum
  factoring algorithm using qubit recycling}.
\newblock \emph{\bibinfo{journal}{Nature Photonics}}
  \textbf{\bibinfo{volume}{6}}, \bibinfo{pages}{773-776}
  (\bibinfo{year}{2012}).

\bibitem{PhysRevLett.76.3228}
\bibinfo{author}{Griffiths, R.~B.} \& \bibinfo{author}{Niu, C.-S.}
\newblock \bibinfo{title}{Semiclassical Fourier transform for quantum
  computation}.
\newblock \emph{\bibinfo{journal}{Phys. Rev. Lett.}}
  \textbf{\bibinfo{volume}{76}}, \bibinfo{pages}{3228-3231}
  (\bibinfo{year}{1996}).

\bibitem{Scie.1110335.10.1126}
\bibinfo{author}{Chiaverini, J.} \emph{et~al.}
\newblock \bibinfo{title}{Implementation of the semiclassical quantum Fourier
  transform in a scalable system}.
\newblock \emph{\bibinfo{journal}{Science}} \textbf{\bibinfo{volume}{308}},
  \bibinfo{pages}{997-1000} (\bibinfo{year}{2005}).

\bibitem{Amico2019}
\bibinfo{author}{Amico, M.}, \bibinfo{author}{Saleem, Z.~H.} \&
  \bibinfo{author}{Kumph, M.}
\newblock \bibinfo{title}{An experimental study of Shor's factoring algorithm
  on ibm q}.
\newblock \emph{\bibinfo{journal}{Phys. Rev. A}}
  \textbf{\bibinfo{volume}{100}}, \bibinfo{pages}{012305}
  (\bibinfo{year}{2019}).

\bibitem{Pal_2019}
\bibinfo{author}{Pal, S.}, \bibinfo{author}{Moitra, S.},
  \bibinfo{author}{Anjusha, V.~S.}, \bibinfo{author}{Kumar, A.} \&
  \bibinfo{author}{Mahesh, T.~S.}
\newblock \bibinfo{title}{Hybrid scheme for factorisation: Factoring 551 using
  a 3-qubit NMR quantum adiabatic processor}.
\newblock \emph{\bibinfo{journal}{Pramana}} \textbf{\bibinfo{volume}{92}},
  \bibinfo{pages}{26} (\bibinfo{year}{2019}).

\bibitem{PhysRevLett.108.130501}
\bibinfo{author}{Xu, N.} \emph{et~al.}
\newblock \bibinfo{title}{Quantum factorization of 143 on a dipolar-coupling
  nuclear magnetic resonance system}.
\newblock \emph{\bibinfo{journal}{Phys. Rev. Lett.}}
  \textbf{\bibinfo{volume}{108}}, \bibinfo{pages}{130501}
  (\bibinfo{year}{2012}).

\bibitem{Saxena_2020}
\bibinfo{author}{Saxena, A.}, \bibinfo{author}{Shukla, A.} \&
  \bibinfo{author}{Pathak, A.} (\bibinfo{year}{2020}).
\newblock \eprint{2009.05840}.

\bibitem{PhysRevLett.85.3049}
\bibinfo{author}{Parker, S.} \& \bibinfo{author}{Plenio, M.~B.}
\newblock \bibinfo{title}{Efficient factorization with a single pure qubit and
  $\mathrm{log}\mathit{N}$ mixed qubits}.
\newblock \emph{\bibinfo{journal}{Phys. Rev. Lett.}}
  \textbf{\bibinfo{volume}{85}}, \bibinfo{pages}{3049-3052}
  (\bibinfo{year}{2000}).

\bibitem{QIC2011517.2011525}
\bibinfo{author}{Beauregard, S.}
\newblock \bibinfo{title}{Circuit for Shor's algorithm using 2n+3 qubits}.
\newblock \emph{\bibinfo{journal}{Quantum Info. Comput.}}
  \textbf{\bibinfo{volume}{3}}, \bibinfo{pages}{175-185}
  (\bibinfo{year}{2003}).

\bibitem{PhysRevA.54.1034}
\bibinfo{author}{Beckman, D.}, \bibinfo{author}{Chari, A.~N.},
  \bibinfo{author}{Devabhaktuni, S.} \& \bibinfo{author}{Preskill, J.}
\newblock \bibinfo{title}{Efficient networks for quantum factoring}.
\newblock \emph{\bibinfo{journal}{Phys. Rev. A}} \textbf{\bibinfo{volume}{54}},
  \bibinfo{pages}{1034-1063} (\bibinfo{year}{1996}).

\bibitem{Marg_1994}
\bibinfo{author}{Margolus, N.}
\newblock \bibinfo{title}{Simple quantum gates}.
\newblock \emph{\bibinfo{journal}{Unpublished manuscript (circa 1994)}}
  (\bibinfo{year}{1994}).

\bibitem{Song_2003}
\bibinfo{author}{Song, G.} \& \bibinfo{author}{Klappenecker, A.}
\newblock \bibinfo{title}{Optimal realizations of simplified Toffoli gates}.
\newblock \emph{\bibinfo{journal}{Quant. Inf. Comp.}} \textbf{\bibinfo{volume}{4}},
  \bibinfo{pages}{361-372} (\bibinfo{year}{2004}).

\bibitem{PhysRevA.93.022311}
\bibinfo{author}{Maslov, D.}
\newblock \bibinfo{title}{Advantages of using relative-phase Toffoli gates with
  an application to multiple control Toffoli optimization}.
\newblock \emph{\bibinfo{journal}{Phys. Rev. A}} \textbf{\bibinfo{volume}{93}},
  \bibinfo{pages}{022311} (\bibinfo{year}{2016}).

\bibitem{Braunstein_1999}
\bibinfo{author}{Braunstein, S.~L.} \emph{et~al.}
\newblock \bibinfo{title}{Separability of very noisy mixed states and
  implications for NMR quantum computing}.
\newblock \emph{\bibinfo{journal}{Phys. Rev. Lett.}}
  \textbf{\bibinfo{volume}{83}}, \bibinfo{pages}{1054-1057}
  (\bibinfo{year}{1999}).

\bibitem{Jozsa_2003}
\bibinfo{author}{Jozsa, R.} \& \bibinfo{author}{Linden, N.}
\newblock \bibinfo{title}{On the role of entanglement in quantum-computational
  speed-up}.
\newblock \emph{\bibinfo{journal}{Proc. Roy. Soc. A}}
  \textbf{\bibinfo{volume}{459}}, \bibinfo{pages}{2011-2032}
  (\bibinfo{year}{2003}).

\bibitem{PhysRevLett.92.087902}
\bibinfo{author}{Bourennane, M.} \emph{et~al.}
\newblock \bibinfo{title}{Experimental detection of multipartite entanglement
  using witness operators}.
\newblock \emph{\bibinfo{journal}{Phys. Rev. Lett.}}
  \textbf{\bibinfo{volume}{92}}, \bibinfo{pages}{087902}
  (\bibinfo{year}{2004}).

\bibitem{TOTH2008430}
\bibinfo{author}{T\'oth, G.}
\newblock \bibinfo{title}{Qubit4matlab v3.0: A program package for quantum
  information science and quantum optics for matlab}.
\newblock \emph{\bibinfo{journal}{Comput. Phys. Commun.}}
  \textbf{\bibinfo{volume}{179}}, \bibinfo{pages}{430-437}
  (\bibinfo{year}{2008}).

\bibitem{PhysRevA.93.032140}
\bibinfo{author}{Ma, X.} \emph{et~al.}
\newblock \bibinfo{title}{Pure-state tomography with the expectation value of
  pauli operators}.
\newblock \emph{\bibinfo{journal}{Phys. Rev. A}} \textbf{\bibinfo{volume}{93}},
  \bibinfo{pages}{032140} (\bibinfo{year}{2016}).

\bibitem{PhysRevLett.105.150401}
\bibinfo{author}{Gross, D.}, \bibinfo{author}{Liu, Y.-K.},
  \bibinfo{author}{Flammia, S.~T.}, \bibinfo{author}{Becker, S.} \&
  \bibinfo{author}{Eisert, J.}
\newblock \bibinfo{title}{Quantum state tomography via compressed sensing}.
\newblock \emph{\bibinfo{journal}{Phys. Rev. Lett.}}
  \textbf{\bibinfo{volume}{105}}, \bibinfo{pages}{150401}
  (\bibinfo{year}{2010}).

\bibitem{Nphyss41567-020-0932-7}
\bibinfo{author}{Huang, H.-Y.}, \bibinfo{author}{Kueng, R.} \&
  \bibinfo{author}{Preskill, J.}
\newblock \bibinfo{title}{Predicting many properties of a quantum system from
  very few measurements}.
\newblock \emph{\bibinfo{journal}{Nature Physics}}
  \textbf{\bibinfo{volume}{16}}, \bibinfo{pages}{1050-1057}
  (\bibinfo{year}{2020}).

\bibitem{PhysRevA.72.022340}
\bibinfo{author}{T\'oth, G.} \& \bibinfo{author}{G\"uhne, O.}
\newblock \bibinfo{title}{Entanglement detection in the stabilizer formalism}.
\newblock \emph{\bibinfo{journal}{Phys. Rev. A}} \textbf{\bibinfo{volume}{72}},
  \bibinfo{pages}{022340} (\bibinfo{year}{2005}).

\bibitem{PhysRevX.7.021042}
\bibinfo{author}{Baccari, F.}, \bibinfo{author}{Cavalcanti, D.},
  \bibinfo{author}{Wittek, P.} \& \bibinfo{author}{Ac\'{\i}n, A.}
\newblock \bibinfo{title}{Efficient device-independent entanglement detection
  for multipartite systems}.
\newblock \emph{\bibinfo{journal}{Phys. Rev. X}} \textbf{\bibinfo{volume}{7}},
  \bibinfo{pages}{021042} (\bibinfo{year}{2017}).

\bibitem{PhysRevLett.117.210504}
\bibinfo{author}{Knips, L.}, \bibinfo{author}{Schwemmer, C.},
  \bibinfo{author}{Klein, N.}, \bibinfo{author}{Wie\ifmmode~\acute{s}\else
  \'{s}\fi{}niak, M.} \& \bibinfo{author}{Weinfurter, H.}
\newblock \bibinfo{title}{Multipartite entanglement detection with minimal
  effort}.
\newblock \emph{\bibinfo{journal}{Phys. Rev. Lett.}}
  \textbf{\bibinfo{volume}{117}}, \bibinfo{pages}{210504}
  (\bibinfo{year}{2016}).

\bibitem{Banaszek_2013}
\bibinfo{author}{Banaszek, K.}, \bibinfo{author}{Cramer, M.} \&
  \bibinfo{author}{Gross, D.}
\newblock \bibinfo{title}{Focus on quantum tomography}.
\newblock \emph{\bibinfo{journal}{New Journal of Physics}}
  \textbf{\bibinfo{volume}{15}}, \bibinfo{pages}{125020}
  (\bibinfo{year}{2013}).

\bibitem{PhysRevA.54.139}
\bibinfo{author}{Barenco, A.}, \bibinfo{author}{Ekert, A.},
  \bibinfo{author}{Suominen, K.-A.} \& \bibinfo{author}{T\"orm\"a, P.}
\newblock \bibinfo{title}{Approximate quantum Fourier transform and
  decoherence}.
\newblock \emph{\bibinfo{journal}{Phys. Rev. A}} \textbf{\bibinfo{volume}{54}},
  \bibinfo{pages}{139--146} (\bibinfo{year}{1996}).

\bibitem{Qiskit}
\bibinfo{author}{H{\'e}ctor Abraham {\it et al.}}
\newblock \bibinfo{title}{Qiskit: An Open-source Framework for Quantum Computing}.
\newblock \emph{\bibinfo{journal}{Available at 10.5281/zenodo.2562110 (2019)}}

\bibitem{PhysRevLett.108.070502}
\bibinfo{author}{Smolin, J. A., Gambetta, J. M. \& Smith, G.}
\newblock \bibinfo{title}{Efficient Method for Computing the Maximum-Likelihood Quantum State from Measurements with Additive Gaussian Noise}.
\newblock \emph{\bibinfo{journal}{Phys. Rev. Lett.}}
  \textbf{\bibinfo{volume}{108}}, \bibinfo{pages}{070502}
  (\bibinfo{year}{2012}).

\bibitem{qmeas_cal}
\bibinfo{author}{Qiskit}
\newblock \bibinfo{title}{Learn quantum computing using Qiskit}.
\newblock \emph{\bibinfo{journal}{Available at https://qiskit.org/textbook/ch-quantum-hardware/measurement-error-mitigation.html}}

\end{thebibliography}

\appendix
\onecolumngrid
\newpage


\section{Postselection scaling \label{sec:postselect}}

In order to do mid-circuit measurements and post select the outcomes, we need to know the basis to measure in for each of the qubits. Thus, one would need to measure qubit $1$ (or the first iteration in the recycling case) in the $\{ \ket{+},\ket{-}\}$ basis, then qubit $2$ (or the second iteration in the recycling case) in either the $\{ S\ket{+},S\ket{-}\}$ or $\{ \ket{+},\ket{-}\}$ basis, then qubit $3$ in either the $\{ TS\ket{+},TS\ket{-}\}$, $\{ S\ket{+},S\ket{-}\}$ or $\{ \ket{+},\ket{-}\}$ basis. Thus, the number of measurements needed scales as $n!$, which grows faster than an exponential with constant base, e.g. $2^n$. So in general the speed up gained would be lost for general factoring using a post selection method, {\it i.e.} factoring numbers larger than $21$.


\section{Effect of relative phase Toffolis \label{sec:effect_of_relative_toffolis}}

Below we show the compiled circuit for the period-finding routine and label specific instances during the evolution of the computation. The aim is to show the invariance of the computation when replacing Toffoli gates with relative phase Toffoli gates that use fewer resources.  
\begin{figure}[hbt]
    \centering
    \mbox{
        \Qcircuit @C=.8em @R=1.0em {
            \lstick{\ket{c_0}}  & \gate{H} & \qw & \qw       & \qw &  \qw     & \qw       & \qw      & \qw
                                &\qw  & \qw      & \ctrl{4}  & \qw      & \qw & \qw       & \ctrl{3} & \qw
                                & \qw      & \multigate{2}{QFT^{\dagger}}     & \meter  & \cw \\
            \lstick{\ket{c_1}} & \gate{H} & \qw & \qw       & \ctrl{3} & \qw & \qw       & \ctrl{2} & \qw
                                & \qw      & \qw       & \qw      & \qw       & \qw      & \qw
                                & \qw      & \qw       & \qw & \ghost{QFT^{\dagger}}& \meter & \cw \\
            \lstick{\ket{c_2}} & \gate{H} & \qw & \ctrl{2}  & \qw      & \qw & \qw       & \qw      & \qw
                                & \qw      & \qw       & \qw      & \qw       & \qw      & \qw
                                & \qw      & \qw       & \qw & \ghost{QFT^{\dagger}} &\meter & \cw \\
            \lstick{\ket{q_0}} & \qw      & \qw  & \qw    & \qw    & \qw   & \targ     & \ctrl{1} & \targ
                                & \qw      & \qw  & \mtarg     & \qw & \qw      & \targ     & \ctrl{1} & \targ
                                & \qw      & \qw & \qw  & \qw \\
            \lstick{\ket{q_1}} & \qw      & \qw & \targ     & \targ    &\qw & \ctrl{-1} & \mtarg    & \ctrl{-1}
                                & \qw      & \gate{X} & \ctrl{-1} & \gate{X}
                                & \qw      & \ctrl{-1} & \mtarg & \ctrl{-1} 
                                & \qw      & \qw       & \qw & \qw \\
                                & & \ustick{\Uparrow} & & & \ustick{\Uparrow}
                                & & & & \ustick{\Uparrow}
                                & & & & \ustick{\Uparrow} & & & & & \\
                                & & \ustick{\ket{\Psi_0}} & & & \ustick{\ket{\Psi_1}}
                                & & & & \ustick{\ket{\Psi_2}}
                                & & & & \ustick{\ket{\Psi_3}} & & & & &
        }
    }
    \caption{States in both registers at various points during the execution of the circuit.}
    \label{fig:checkpoints}
\end{figure}
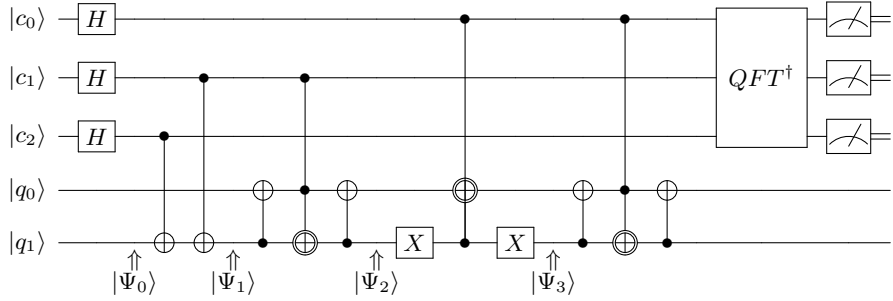

The states are various points of the evolution are given explicitly as
\begin{align}
    \ket{\Psi_0} &= \ket{+}_{c_0}\ket{+}_{c_1}\ket{+}_{c_2}\ket{0}_{q_0}\ket{0}_{q_1}, \nonumber \\ \nonumber \\
    \ket{\Psi_1} &= \ket{+}_{c_0}(\ket{0}_{c_1}\ket{0}_{c_2}\ket{0}_{q_0}\ket{0}_{q_1} + \ket{0}_{c_1}\ket{1}_{c_2}\ket{0}_{q_0}\ket{1}_{q_1} + \nonumber \\ 
                 &\>\qquad\qquad \ket{1}_{c_1}\ket{0}_{c_2}\ket{0}_{q_0}\ket{1}_{q_1} + \ket{1}_{c_1}\ket{1}_{c_2}\ket{0}_{q_0}\ket{0}_{q_1}), \nonumber \\ \nonumber \\
    \ket{\Psi_2} &= \ket{0}_{c_0}\ket{0}_{c_1}\ket{0}_{c_2}\ket{0}_{q_0}\ket{0}_{q_1} + \ket{0}_{c_0}\ket{0}_{c_1}\ket{1}_{c_2}\ket{0}_{q_0}\ket{1}_{q_1} + \nonumber \\
                 &\>\>\quad \ket{0}_{c_0}\ket{1}_{c_1}\ket{0}_{c_2}\ket{1}_{q_0}\ket{0}_{q_1} + \ket{0}_{c_0}\ket{1}_{c_1}\ket{1}_{c_2}\ket{0}_{q_0}\ket{0}_{q_1} + \nonumber \\
                 &\>\>\quad \ket{1}_{c_0}\ket{0}_{c_1}\ket{0}_{c_2}\ket{0}_{q_0}\ket{0}_{q_1} + \ket{1}_{c_0}\ket{0}_{c_1}\ket{1}_{c_2}\ket{0}_{q_0}\ket{1}_{q_1} + \nonumber \\
                 &\>\>\quad \ket{1}_{c_0}\ket{1}_{c_1}\ket{0}_{c_2}\ket{1}_{q_0}\ket{0}_{q_1} + \ket{1}_{c_0}\ket{1}_{c_1}\ket{1}_{c_2}\ket{0}_{q_0}\ket{0}_{q_1}, \nonumber \\ \nonumber \\
    \ket{\Psi_3} &= \ket{0}_{c_0}\ket{0}_{c_1}\ket{0}_{c_2}\ket{0}_{q_0}\ket{0}_{q_1} + \ket{0}_{c_0}\ket{0}_{c_1}\ket{1}_{c_2}\ket{0}_{q_0}\ket{1}_{q_1} + \nonumber \\
                 &\>\>\quad\ket{0}_{c_0} \ket{1}_{c_1}\ket{0}_{c_2}\ket{1}_{q_0}\ket{0}_{q_1} + \ket{0}_{c_0}\ket{1}_{c_1}\ket{1}_{c_2}\ket{0}_{q_0}\ket{0}_{q_1} + \nonumber \\
                 &\>\>\quad\ket{1}_{c_0}\ket{0}_{c_1}\ket{0}_{c_2}\ket{1}_{q_0}\ket{0}_{q_1} + \ket{1}_{c_0}\ket{0}_{c_1}\ket{1}_{c_2}\ket{0}_{q_0}\ket{1}_{q_1} + \nonumber \\ 
                 &\>\>\quad\ket{1}_{c_0}\ket{1}_{c_1}\ket{0}_{c_2}\ket{0}_{q_0}\ket{0}_{q_1} + \ket{1}_{c_0}\ket{1}_{c_1}\ket{1}_{c_2}\ket{1}_{q_0}\ket{0}_{q_1}.
\end{align}

Looking at the state $\ket{\Psi_1}$, one can see that none of its constituent states is transformed into $\ket{1}_{c_1}\ket{0}_{q_0}\ket{1}_{q_1}$ by the CX gate that follows, since the state $\ket{1}_{c_1}\ket{1}_{q_0}\ket{1}_{q_1}$ that would be transformed to the former is not present in $\ket{\Psi_1}$. Thus the relative phase Toffoli gate does not affect the phase in the registers. 

Similarly for $\ket{\Psi_2}$, the state $\ket{1}_{c_0}\ket{1}_{q_0}\ket{0}_{q_1}$ is not present when the subsequent relative phase Toffoli gate is applied because the state $\ket{1}_{c_0}\ket{1}_{q_0}\ket{1}_{q_1}$ is absent from the register for $\ket{\Psi_2}$ and this is needed when $\hat{X}$ is applied to qubit $q_1$. 

The scenario for $\ket{\Psi_3}$ is the same as that of $\ket{\Psi_1}$, the only difference is the control is now $c_0$. 

The Margolus gates in this particular quantum circuit never encounter the basis state $\ket{101}$, thus the operation of the circuit remains unchanged by the replacement of full Toffoli gates with their respective relative phase counterparts.


\section{IBM Quantum Experience \label{sec:ibm_q_experience}}
The experiments in this paper were conducted on the IBM Quantum Experience \textbf{ibmq\_toronto} and \textbf{ibmq\_casablanca} processors through the software development kit Qiskit~\cite{Qiskit}. Each experiment reported here was conducted on the date shown in the table below.

\begin{table}[hbt]
  \centering
  \fontfamily{ppl}\selectfont
  \begin{tabular}{ll}
    \toprule
    Experiment & Date \\
    \midrule
    Compiled quantum order-finding on \textbf{ibmq\_casablanca} & 2020/12/03 \\
    State tomography on \textbf{ibmq\_casablanca}                & 2020/12/04 \\
    Verification of entanglement on \textbf{ibmq\_casablanca}    & 2020/12/04 \\
    Compiled quantum order-finding on \textbf{ibmq\_toronto}    & 2020/12/06 \\
    Verification of entanglement on \textbf{ibmq\_toronto}       & 2020/12/07 \\
    State tomography on \textbf{ibmq\_toronto}                   & 2020/12/16 \\
    \bottomrule
  \end{tabular}
  \caption{
    Dates of experiments.
  }
  \label{tab:exps_dates}
\end{table}

For characterization purposes, the compiled quantum order-finding experiments were submitted in batches of $900$ circuits with each circuit having 8192 measurement shots. In total, $900\times8192$ measurements were made. In choosing the qubit device mappings shown in the main paper, preference was given to the qubit pairs with relatively small \cx error rates. Tables \ref{tab:errors_toronto} and \ref{tab:errors_casablanca} show reported single qubit-error rates for \textbf{ibmq\_toronto} and \textbf{ibmq\_casablanca} respectively, where $U2(\phi, \lambda) = R_z(\phi)R_y(\frac{\pi}{2})R_z(\lambda)$. Table \ref{tab:cx_errors} shows the \cx error rates for the two processors. The dates of the experiments are given in the captions.

\begin{table}[hbt]
  \centering
  \fontfamily{ppl}\selectfont
  \begin{tabular}{lll}
    \toprule
         & \uu gate error rate & Readout error rate \\
    \midrule
      Q0 & \SI{6.010e-2}{} & \SI{4.39e-4}{}  \\
      Q1 & \SI{3.14e-2}{}  & \SI{2.12e-4}{}  \\
      Q2 & \SI{2.98e-2}{}  & \SI{1.96e-4}{}  \\
      Q3 & \SI{9.30e-3}{}  & \SI{5.74e-4}{}  \\
      Q4 & \SI{1.34e-2}{}  & \SI{2.097e-4}{} \\
    \bottomrule
  \end{tabular}
  \caption{
    Reported single-qubit gate errors on 16 December 2020. 
  }
  \label{tab:errors_toronto}
\end{table}

\begin{table}[hbt]
  \centering
  \fontfamily{ppl}\selectfont
  \begin{tabular}{lll}
    \toprule
    & \uu gate error rate & Readout error rate \\
    \midrule
      Q0 & \SI{2.16e-2}{}  & \SI{2.18e-4}{}  \\
      Q1 & \SI{1.31e-2}{}  & \SI{4.042e-4}{} \\
      Q2 & \SI{1.54e-2}{}  & \SI{2.78e-4}{}  \\
      Q3 & \SI{9.30e-2}{}  & \SI{2.62e-4}{}  \\
      Q4 & \SI{1.67e-2}{}  & \SI{4.96e-4}{}  \\
    \bottomrule
  \end{tabular}
  \caption{
    Reported single-qubit gate errors on 06 December 2020. 
  }
  \label{tab:errors_casablanca}
\end{table}

\begin{table}[hbt]
  \centering
  \fontfamily{ppl}\selectfont
  \begin{tabular}{lll}
    \toprule
    & \textbf{ibmq\_toronto} & \textbf{ibmq\_casablanca} \\
    \midrule
      \cx(0,1) & \SI{6.620e-3}{}  & \SI{9.126e-3}{}  \\
      \cx(1,4) & \SI{8.214e-3}{}  & \SI{1.114e-2}{}  \\
      \cx(2,1) & \SI{7.152e-3}{}  & \SI{7.446e-3}{}  \\
      \cx(3,2) & \SI{6.824e-3}{}  & \SI{1.337e-2}{}  \\
    \bottomrule
  \end{tabular}
  \caption{
    Reported \cx gate errors on 06 December (\textbf{ibmq\_casablanca}) and 16 December (\textbf{ibmq\_toronto}) 2020.
  }
  \label{tab:cx_errors}
\end{table}

Qiskit's state tomography fitter uses a least-squares fitting to find the closest density matrix described by Pauli measurement results~\cite{PhysRevLett.108.070502}. On an $n$-qubit system, the fitter requires measurement results from executing $3^n$ circuits. This makes state tomography on large circuits in impractical. Thus only $30$ state tomography experiments were performed for the three control register qubits and in total $3^3 \times 30 \times 8192$ measurement were made.

In reducing the effect of noise due to final measurement errors, Qiskit recommends a measurement error mitigation approach. The approach starts off by creating circuits that each perform a measurement of the $2^n$ basis states. The measurement counts of the $2^n$ basis state measurements are put into a column vector $C_{noisy}$, arranged in ascending order by the value of their measurement bitstring, {\it i.e.} $00\ldots00$ is the first element, the next is $00\ldots01$ and so on. The approach assumes that there is a matrix $M$ called the calibration matrix, such that
\begin{align}
  C_\text{noisy} &= M C_\text{ideal},
\end{align}
where $C_\text{ideal}$ is a column vector of measurement counts in the absence of noise. If $M$ is invertible then, then $C_\text{noisy}$  can transformed into $C_\text{ideal}$ by finding $M^{-1}$
\begin{align}
  C_\text{ideal} &=  M^{-1}C_\text{noisy}.
\end{align}
Qiskit~\cite{qmeas_cal} uses a least-squares fit to calculate an approximate $M^{-1}$ by some other matrix $\tilde{M}^{-1}$, as in general $M$ is not invertible, giving
\begin{align}
  C_\text{mitigated} &=  \tilde{M}^{-1}C_\text{noisy}.
\end{align}

The entries of the column vector $C_\text{mitigated}$ correspond to the mitigated measurement counts in same order as before. The entirety of the results reported in our work make of use of this approach.


\section{Error bars \label{sec:error_bars}}

All the confidence intervals of the data presented here were established via non-parametric bootstrap resampling techniques. In order to place the constraint that the measurement counts should sum to the number of experimental shots, a sample contains data as column vectors of outcomes of some experiment. In each round, the resampling draws entire column vectors whose elements respect the aforementioned constraint. For each outcome across the column vectors, mean estimates are obtained and a confidence interval around the estimates can be appropriately constructed. 

To elucidate the above, consider the following example. Consider the outcomes of a two-qubit experiment with experimental shots of $8192$ repeated $4$ times, as shown in Table~\ref{tab:bootstrap_example} below.
\begin{table}[hbt]
  \centering
  \fontfamily{ppl}\selectfont
  \begin{tabular}{lllll}
    \toprule
    Outcomes & \multicolumn{3}{c}{Counts} \\
    \midrule
     & Exp. 1 & Exp. 2 & Exp. 3 & Exp. 4 \\
    \midrule
    00 & 2335 & 2208 & 2406 & 2203\\ 
    01 & 665  & 690  & 633  & 656 \\ 
    10 & 183  & 100  & 197  & 177 \\
    11 & 5009 & 5192 & 4956 & 5156 \\
    \bottomrule
  \end{tabular}
  \caption{
      Example data for a two-qubit experiment repeated $4$ times for illustrating how bootstrap resampling was done.
  }
  \label{tab:bootstrap_example}
\end{table}

Suppose we resampled the experiments $1,1,2,4$ from Table~\ref{tab:bootstrap_example}, making a bootstrap sample of size $5$.
\begin{align}
    B = [[2335, 665, 183, 5009], \nonumber \\
         [2335, 665, 183, 5009], \nonumber \\
         [2208, 690, 100, 5192],\nonumber \\
         [2203, 656, 177, 5156]].
\end{align}
From this, we can obtain appropriately the bootstrap sample for each outcome (corresponding to an index), {\it e.g.} the bootstrap sample for the outcomes at index $0$ (outcome 00) is
\begin{align}
    B_{0} = [2335, 2335, 2208, 2203].
\end{align}
The bootstrap mean estimates and confidence intervals can then be performed for each outcome while respecting the constraint of the measurement counts summing up to the total number of experimental shots.


\section{Pauli measurements \label{sec:pauli_measurements}}
As an example, consider the measurement of the Pauli expectation value $\expval{ZZZZZ}$. Let $p_{ijklm}$ denote the probability for a computational basis measurement $\{\ket{0}, \ket{1}\}$ of five qubits to output the binary string $ijklm$, {\it i.e.} $p_{00000}$ denotes the probability to measure all the qubits in $\ket{0}$ state. To calculate $\expval{ZZZZZ}$ we can combine these probabilities as given in the equation below
\begin{align}
    \label{eq:zzzzz}
    \expval{ZZZZZ} &= p_{00000} - p_{00010} - p_{00100} + p_{00101} + p_{00110} - p_{01000} + p_{01001} + p_{01010} + p_{01100} - p_{01101} - \nonumber \\
                   & \quad\> p_{01110} + p_{01111} - p_{10000} + p_{10001} + p_{10010} - p_{10011} + p_{10100} - p_{10101} - p_{10110} + p_{10111} + \nonumber \\
                   & \quad\> p_{11000} - p_{11001} - p_{11010} + p_{11011} - p_{11100} + p_{11101} + p_{11110} - p_{11111}.
\end{align}
Similarly, the expectation $\expval{IZIZI}$ is given by
\begin{align}
    \label{eq:izizi}
    \expval{IZIZI} &= p_{00000} - p_{00010} + p_{00100} + p_{00101} - p_{00110} - p_{01000} - p_{01001} + p_{01010} - p_{01100} - p_{01101} + \nonumber \\
                   &\quad\>  p_{01110} + p_{01111} + p_{10000} + p_{10001} - p_{10010} - p_{10011} + p_{10100} + p_{10101} - p_{10110} - p_{10111} - \nonumber \\
                   &\quad\>  p_{11000} - p_{11001} + p_{11010} + p_{11011} - p_{11100} - p_{11101} + p_{11110} + p_{11111}.
\end{align}
However, the terms in the equation above are given by the marginalization of the distribution measured in Eq.~\eqref{eq:zzzzz} across the outcome space of qubits $1$, $3$ and $5$. By considering all such marginalizations of the distribution in Eq.~\eqref{eq:zzzzz}, we obtain the set of Pauli expectation values that can be derived from a measurement of $\expval{ZZZZZ}$, namely
\begin{align}
  \{\> & ZZZZI, ZZZIZ, ZZZII, ZZIZZ, ZZIZI, ZZIIZ, ZZIII, ZIZZZ, ZIZZI, ZIZIZ, ZIZII, ZIIZZ, \nonumber \\
     & ZIIZI, ZIIIZ, ZIIII, IZZZZ, IZZZI, IZZIZ, IZZII, IZIZZ, IZIZI, IZIIZ, IZIII, IIZZZ, \nonumber \\
     & IIZZI, IIZIZ, IIZII, IIIZZ, IIIZI, IIIIZ \>\}.
\end{align}
After applying what is described above to the Pauli decomposition of the ideal state $\rho=\ketbra{\Psi}{\Psi}$, we reduce the number of terms that we need to measure from $293$ to $79$ terms, as given below
\begin{align}
\{\> & XXXXZ, XXXZX, XXXZZ, XXYYZ, XXYZY, XXZXX, XXZXZ, XXZYY, XXZZX, XYXYZ, \nonumber \\
  &  XYXZY, XYYXZ, XYYZX, XYYZZ, XYZXY, XYZYX, XYZYZ, XYZZY, XZXXX, XZXYY, \nonumber \\
  &  XZXZZ, XZYXY, XZYYX, XZZXZ, XZZZX, YXXYZ, YXXZY, YXYXZ, YXYZX, YXYZZ, \nonumber \\
  &  YXZXY, YXZYX, YXZYZ, YXZZY, YYXXZ, YYXZX, YYXZZ, YYYYZ, YYYZY, YYZXX, \nonumber \\
  &  YYZXZ, YYZYY, YYZZX, YZXXY, YZXYX, YZYXX, YZYYY, YZYZZ, YZZYZ, YZZZY, \nonumber \\
  &  ZXXXZ, ZXXZX, ZXXZZ, ZXYYZ, ZXYZY, ZXZXX, ZXZXZ, ZXZYY, ZXZZX, ZYXYZ, \nonumber \\
  &  ZYXZY, ZYYXZ, ZYYZX, ZYYZZ, ZYZXY, ZYZYX, ZYZYZ, ZYZZY, ZZXXX, ZZXXZ, \nonumber \\
  &  ZZXYY, ZZXZX, ZZYXY, ZZYYX, ZZYYZ, ZZYZY, ZZZXX, ZZZYY, ZZZZZ \>\}.
\end{align}


\section{Maximum overlap with respect to the bipartitions \label{sec:maximum_overlap_wrt_bipartitions}}
The values listed below were obtained using the software package QUBIT4MATLAB~\cite{TOTH2008430}. Here, $\ket{\phi}$ is a pure biseparable state in some defined bipartite partition (bipartition), \emph{i.e.} an unentangled product state with respect to this bipartition, and $\ket{\Psi}$ is the ideal state in both the control and work registers preceding the application of the QFT to the control register.
\begin{align}
    \underset{\phi \in \{(c_1)(c_0c_2q_0q_1)\}}{\max}\abs{\braket{ \phi}{\Psi }}^{2} &= 0.500, \nonumber \\
    \underset{\phi \in \{(c_2)(c_0c_1q_0q_1)\}}{\max}\abs{\braket{ \phi}{\Psi }}^{2} &= 0.500, \nonumber \\
    \underset{\phi \in \{(q_0)(c_0c_1c_2q_1)\}}{\max}\abs{\braket{ \phi}{\Psi }}^{2} &= 0.750, \nonumber \\
    \underset{\phi \in \{(q_1)(c_0c_1c_2q_0)\}}{\max}\abs{\braket{ \phi}{\Psi }}^{2} &= 0.625, \nonumber \\
    \underset{\phi \in \{(c_0c_1)(c_2q_0q_1)\}}{\max}\abs{\braket{ \phi}{\Psi }}^{2} &= 0.500, \nonumber \\
    \underset{\phi \in \{(c_0c_2)(c_1q_0q_1)\}}{\max}\abs{\braket{\phi}{\Psi }}^{2} &= 0.500, \nonumber \\
    \underset{\phi \in \{(c_0q_0)(c_1c_2q_1)\}}{\max}\abs{\braket{ \phi}{\Psi }}^{2} &= 0.427, \nonumber \\
    \underset{\phi \in \{(c_0q_1)(c_1c_2q_0)\}}{\max}\abs{\braket{ \phi}{\Psi }}^{2} &= 0.570, \nonumber \\
    \underset{\phi \in \{(q_0q_1)(c_0c_1c_2)\}}{\max}\abs{\braket{ \phi}{\Psi }}^{2} &= 0.375, \nonumber \\
    \underset{\phi \in \{(c_0q_1)(c_0c_1q_0)\}}{\max}\abs{\braket{ \phi}{\Psi }}^{2} &= 0.570, \nonumber \\
    \underset{\phi \in \{(c_1q_1)(c_0c_2q_0)\}}{\max}\abs{\braket{ \phi}{\Psi }}^{2} &= 0.570, \nonumber \\
    \underset{\phi \in \{(c_1q_0)(c_0c_2q_1)\}}{\max}\abs{\braket{ \phi}{\Psi }}^{2} &= 0.427, \nonumber \\
    \underset{\phi \in \{(c_2q_0)(c_0c_1q_1)\}}{\max}\abs{\braket{ \phi}{\Psi }}^{2} &= 0.427, \nonumber \\
    \underset{\phi \in \{(c_1c_2)(c_0q_0q_1)\}}{\max}\abs{\braket{ \phi}{\Psi }}^{2} &= 0.500.
\end{align}
For a given separation of the qubits into two partitions (a bipartition), {\it e.g.} $(c_1)(c_0c_2q_0q_1)$, there is a pure product state $\ket{\phi}$ with respect to these partitions, {\it i.e.} no entanglement between the partitions, that maximizes the overlap squared with the ideal state. The value of the overlap squared between this product state and the ideal state, {\it e.g.} 0.5, is therefore the highest value that can be obtained for an unentangled state between the partitions. Thus, if a given state has an overlap squared larger than 0.5 it must be an entangled state with respect to the partitions. The value of the maximum overlap squared changes for the different partitions chosen as it depends on the structure of the ideal state. The above results extend to mixed states across the bipartitions due to the convex sum nature of quantum states~\cite{TOTH2008430}.


\section{Continued fractions and convergents}\label{sec:continued_fractions}
A $2L + 1$ bit rational number $\varphi$ is said to have a continued fraction expansion if it can be written as
\begin{align}
    \varphi \equiv [a_0, a_1, \ldots, a_n] \equiv a_0 + \frac{1}{a_1 + \frac{1}{a_2 +  \frac{1}{\cdots + \frac{1}{a_n}}}},
\end{align}
where $n$ is a finite integer and the $a_i$'s are integers. Additionally, if $\varphi < 1$, we have $a_0 = 0$. The convergents of the continued fraction expansion are the rationals,

\begin{align}\label{eq:convergents}
    a_0, a_0 + \frac{1}{a_1}, a_0 + \frac{1}{a_1 + \frac{1}{a_2}}, \cdots
\end{align}

If a rational number $s/r$ satisfies the following inequality
\begin{align}
    \abs{\frac{s}{r} - \varphi} \leq \frac{1}{2r^2},
\end{align}
then $s/r$ will appear as a convergent in the continued fraction expansion of $\varphi$. If $\varphi$ is an approximation of $s/r$ accurate to $2L + 1$ bits, then we have $\abs{s/r - \varphi} \leq 1/2^{2L + 1}$. For $r \leq N \leq 2^L$, we have that $1/2^{2L + 1} \leq 1/2r^2$. Therefore, since the inequality holds for the approximation $\varphi$, there is a classical algorithm that can compute the convergents of $\varphi$, and produce integers $s', r'$ such that $\text{gcd}(s', r') = 1$ in $\bigO{L^3}$ operations~\cite{Mike&Ike}. We can then check if $r'$ is the order of $a$ and $N$ by testing whether $a^{r'}\>\text{mod}\>N = 1$. Note that in our approach, $\varphi=\varphi_s/2^n\simeq s/r$ is not an approximation that is accurate to $2L+1$ bits as above, but is a further approximation of $s/r$ depending on the resolution, {\it i.e.} the number of iterations, or alternatively qubits in the control register.

Consider the following example of the final measurement outcomes from Fig.~7 in the main text, where the outcome $\ket{110} = \ket{6}$ is not a peak but $\ket{101} = \ket{5}$ is a peak in the outcome distribution and we have used the integer representation of the binary outcome. The former outcome gives $\varphi = \frac{6}{2^3}$ and latter gives $\varphi = \frac{5}{2^3}$. Computing the continued fractions of the former gives
\begin{align}
    \frac{6}{8} &= \frac{3}{4}, \nonumber \\
    \frac{3}{4} &= 0 + \frac{1}{\frac{4}{3}}, \nonumber \\
    \frac{3}{4} &= 0 + \frac{1}{1 + \frac{1}{3}}.
\end{align}
Thus
\begin{align}
    \frac{6}{8} = [0, 1, 3].
\end{align}
Computing the convergents according to Eq.~\eqref{eq:convergents} gives $0, 1, 3/4$. 

On the other hand, computing the continued fractions of the latter $\varphi$ gives
\begin{align}
    \frac{5}{8} &= 0 + \frac{1}{\frac{8}{5}}, \nonumber \\
    \frac{5}{8} &= 0 + \frac{1}{1 + \frac{3}{5}}, \nonumber \\
    \frac{5}{8} &= 0 + \frac{1}{1 + \frac{1}{\frac{5}{3}}}, \nonumber \\
    \frac{5}{8} &= 0 + \frac{1}{1 + \frac{1}{1 + \frac{2}{3}}}, \nonumber \\
    \frac{5}{8} &= 0 + \frac{1}{1 + \frac{1}{1 + \frac{1}{\frac{3}{2}}}}, \nonumber \\
    \frac{5}{8} &= 0 + \frac{1}{1 + \frac{1}{1 + \frac{1}{1 + \frac{1}{2}}}}.
\end{align}
This gives 
\begin{align}
    \frac{5}{8} = [0, 1, 1, 1, 2].
\end{align}
Computing the convergents gives $0, 1, 1/2, 2/3, 5/8$. Looking at the former and latter computed convergents, we note that the third convergent of the latter correctly gives $r' = 3$ while the convergents of the former do not give the correct order when tested using $a^{r'}\>\text{mod}\>N = 1$. The same process can be applied to the outcome $\ket{011} = \ket{3}$, which is a peak and correctly gives $r'=3$.

\end{document}